\documentclass[a4paper,aps,prx,longbibliography,floatfix,showpacs,superscriptaddress]{revtex4-1}

\usepackage{times}
\usepackage{verbatim}
\usepackage{bbold}
\usepackage{bbm}
\usepackage[pdftex]{graphicx}
\usepackage{latexsym,amsmath,verbatim}
\usepackage{color}
\usepackage{hyperref}
\usepackage{lipsum}
\usepackage{rotating}
\usepackage{multirow}
\usepackage{microtype}
\usepackage{comment}
\usepackage[normalem]{ulem}
\usepackage{longtable}

\newcommand{\av}[1]{\left\langle {#1} \right\rangle}
\newcommand{\km}{q_\mathrm{max}}
\newcommand{\avK}{\av{q}_{K_M}}
\newcommand{\AV}{\av{q^2}/\av{q}}
\newcommand{\CLV}{\av{q^2}/[\av{q} \sqrt{q_\mathrm{max}}]}

\begin{document}

\title{Relating Topological Determinants of Complex Networks to Their
  Spectral Properties: Structural and Dynamical Effects}

\author{Claudio Castellano}

\affiliation{Istituto dei Sistemi Complessi (ISC-CNR), Via dei Taurini 
19, I-00185 Roma, Italy}

\author{Romualdo Pastor-Satorras} \email{Corresponding author:
  romualdo.pastor@upc.edu}

\affiliation{Departament de F\'{\i}sica, Universitat Polit\`ecnica de
  Catalunya, Campus Nord B4, 08034 Barcelona, Spain}

\begin{abstract}
  The largest eigenvalue of a network's adjacency matrix and its
  associated principal eigenvector are key elements for determining the
  topological structure and the properties of dynamical processes
  mediated by it.  We present a physically grounded expression relating
  the value of the largest eigenvalue of a given network to the
  largest eigenvalue of two network subgraphs, considered as isolated:
  The hub with its immediate neighbors and the densely connected set of
  nodes with maximum $K$-core index.  We validate this formula showing
  that it predicts with good accuracy the largest eigenvalue of a large
  set of synthetic and real-world topologies.  We also present evidence 
  of the consequences of these findings for broad
  classes of dynamics taking place on the networks. As a byproduct, we
  reveal that the spectral properties of heterogeneous networks built
  according to the linear preferential attachment model are
  qualitatively different from those of their static counterparts.
\end{abstract}


\maketitle

\section{Introduction}
\label{sec:introduction}

The spectral properties of complex topologies~\cite{PVM_graphspectra}
play a crucial role in our understanding of the structure and function
of real networked systems. Various matrices can be constructed for any
given network, their spectral properties accounting for different
topological or functional features. Thus, for example, the Laplacian
matrix is related to diffusive and random walk dynamics on
networks~\cite{Samukhin2008}, the modularity matrix plays a role in
community identification on networks~\cite{Fortunato201075}, while the
non-backtracking or Hashimoto matrix governs
percolation~\cite{Karrer2014}. Among all matrices associated to
networks, the simplest and possibly most studied is the adjacency matrix
$A_{ij}$, taking the value $1$ whenever nodes $i$ and $j$ are connected,
and zero otherwise. Particular interest in this case is placed on the
study of the principal eigenvector $\{ f_i\}$ (PEV), defined as the
eigenvector of the adjacency matrix with the largest eigenvalue
$\Lambda_M$ (LEV). This interest is twofold.  On the one hand, the PEV
is one of the fundamental measures of node importance or
centrality~\cite{Newman10}.  The centrality of a node can be defined
based on the number of other different vertices that can be reached from
it, or the role it plays in connecting together different parts of the
network. From a more sociological point of view, a node is central if it
is connected to other central nodes. From this definition arises the
notion of eigenvector centrality of a node~\cite{Bonacich72}, that
coincides with the corresponding component of the PEV. On the other
hand, the LEV plays a pivotal role in the behavior of many dynamical
systems on complex networks, such as epidemic
spreading~\cite{Chakrabarti_2008}, synchronization of weakly coupled
oscillators~\cite{Restrepo2005}, weighted percolation on directed
networks~\cite{Restrepo2008}, models of genetic
control~\cite{Pomerance2009}, or the dynamics of excitable
elements~\cite{Kinouchi:2006aa}. In this kind of dynamical processes,
the LEV is related, through different analytical techniques, to the
critical point at which a transition between different phases takes
place: In terms of some generic control parameter $\lambda$, a critical
point $\lambda_c$ is found to be in general inversely proportional to
the LEV $\Lambda_M$.

The possibility to know the position of such transition points in terms
of simple network topological properties is of great importance, as it
allows to predict the system macroscopic behavior or optimize the
network to control processes on it.  This has triggered an intense
activity~\cite{Dorogovtsev2003,Kim2007,Nadakuditi2013,Restrepo07},
particularly in the case of networks with heterogeneous topology, such
as power-law distributed networks with a degree distribution of the form
$P(q) \sim q^{-\gamma}$~\cite{Barabasi:1999}. Among these efforts, in
their seminal work~\cite{Chung03}, Chung, Lu and Vu (CLV) have
rigorously proven, for a model with power-law degree distribution, that
the LEV can be expressed in terms of the maximum degree $\km$ present in
the network and the first two moments of the degree distribution. 
This is a remarkable achievement, as it allows to draw predictions in the
analysis of dynamics on networks.

Here we show that while the CLV theory provides in some cases an
excellent approximation to the LEV, specially in the case of random
uncorrelated networks, it can fail considerably in other cases.  In
order to provide better estimates, we reinterpret the CLV result in
terms of the competition among different subgraphs in the networks. This
insight leads us to the formulation of a modified form of the CLV
theory, that captures the behavior of the LEV more generally, including
the case of real correlated networks, and asymptotically reduces to the
CLV form in the case of random uncorrelated networks.  We show that our
generalized expression perfectly predicts the LEV for linear
preferential attachment growing networks (for which the original CLV
form fails) and provides an excellent approximation for the LEV of
real-world networks.  Finally, we show that our modified expression
predicts reliably the critical point of dynamical processes on a large
set of synthetic and real-world networks, with no exception.

The paper is organized as follows: In Sec.~\ref{sec:chung-lu-vu} we
review the original expression for the largest eigenvalue from
Ref.~\cite{Chung03}, and show how it can lead to large errors in real
correlated networks. In Sec.~\ref{sec:gener-scal-larg} we present
physical arguments substantiating a new generalized expression for the
largest eigenvalue, whose validity is checked against a large set of
real networks. In Sec.~\ref{sec:case-line-pref} we discuss in detail the
case of growing linear preferential attachment networks, which turn out
to be a remarkable benchmark for the plausibility of our new generalized
expression. We discuss the effects of our prediction in the estimation
of the critical point in epidemics and synchronization dynamics in
Sec.~\ref{sec:cons-dynam-netw}. Finally, we present our conclusions and
future avenues of work in Sec.~\ref{sec:Discussion}. Several appendices
provide details and additional information.

\section{The Chung-Lu-Vu formula for the largest eigenvalue}
\label{sec:chung-lu-vu}

In Ref.~\cite{Chung03}, the authors consider a class of network models
with given expected degree distribution.  That is, starting from a predefined
degree distribution $P(q)$, one generates expected degrees $\tilde{q}_i$
for each node, drawn from $P(q)$, and creates an actual network by
joining every pair of nodes $i$ and $j$ with probability
$\tilde{q}_i \tilde{q}_j / \sum_r \tilde{q}_r$. The resulting network
has a degree distribution with the same functional form as the imposed
$P(q)$ and lacks degree correlations, since the condition
$\tilde{q}_i^2 < \sum_r \tilde{q}_r$ is imposed in the
construction~\cite{Chung03,mariancutofss}. This algorithm is a variation
of the classical configuration model~\cite{Nadakuditi2013}, cast in
terms of a hidden variables model~\cite{Boguna2003}. For this model
network, and any arbitrary degree distribution, the authors in
Ref.~\cite{Chung03} rigorously prove that the largest eigenvalue of the
corresponding adjacency matrix takes the form (see
also~\cite{Chung2003AnnComb})
\begin{equation}
  \Lambda_M =
  \left \{ 
    \begin{array}{lcr}
      a_1 \sqrt{\km} &\;\mathrm{if}\;\;& \sqrt{\km} > \frac{\av{q^2}}{\av{q}}
      \ln^2(N) \\ 
      a_2 \frac{\av{q^2}}{\av{q}} &\;\mathrm{if}\;\;& \frac{\av{q^2}}{\av{q}} >
      \sqrt{\km} \ln(N) 
    \end{array}
  \right. ,
  \label{Lambda_M}
\end{equation}
where $N$ is the network size, $\km$ is the maximum degree in the
networks, and $a_i$ are constants of order $1$. In the case of
scale-free networks, the maximum degree is a growing function of $N$,
that for uncorrelated networks~\cite{Dorogovtsev:2002} takes the value
$\km \sim N^{1/2}$ for $\gamma \leq 3$ and $\km \sim N^{1/(\gamma-1)}$
for $\gamma >3$~\cite{mariancutofss}. The algebraic increase of $\km$
allows to disregard the logarithmic terms in Eq.~(\ref{Lambda_M}) in the
limite of infinite size networks, leading to the simpler expression
\cite{Castellano2010}
\begin{equation}
  \label{eq:1}
  \Lambda_M \approx \max \{ \sqrt{\km}, \av{q^2}/\av{q}\},
\end{equation}
valid for any value of $\gamma$. 
For power law distributed networks,
the second moment of the degree distribution scales as
$\av{q^2} \sim \km^{3-\gamma}$ for $\gamma \leq 3$ and
$\av{q^2} \sim \mathrm{const.}$ for $\gamma > 3$. Combining this result
with the expression for the maximum degree, we can write the more
explicit result
\begin{equation}
\label{eq:4}
\Lambda_M \approx \left \{ 
    \begin{array}{lcr}
      \sqrt{\km} &\;\mathrm{if}\;\;& \gamma > 5/2 \\ 
      \frac{\av{q^2}}{\av{q}} &\;\mathrm{if}\;\;&  \gamma < 5/2
    \end{array}
  \right. .
\end{equation}

\begin{figure}[t]
  \centering
  \includegraphics[width=0.8\textwidth]{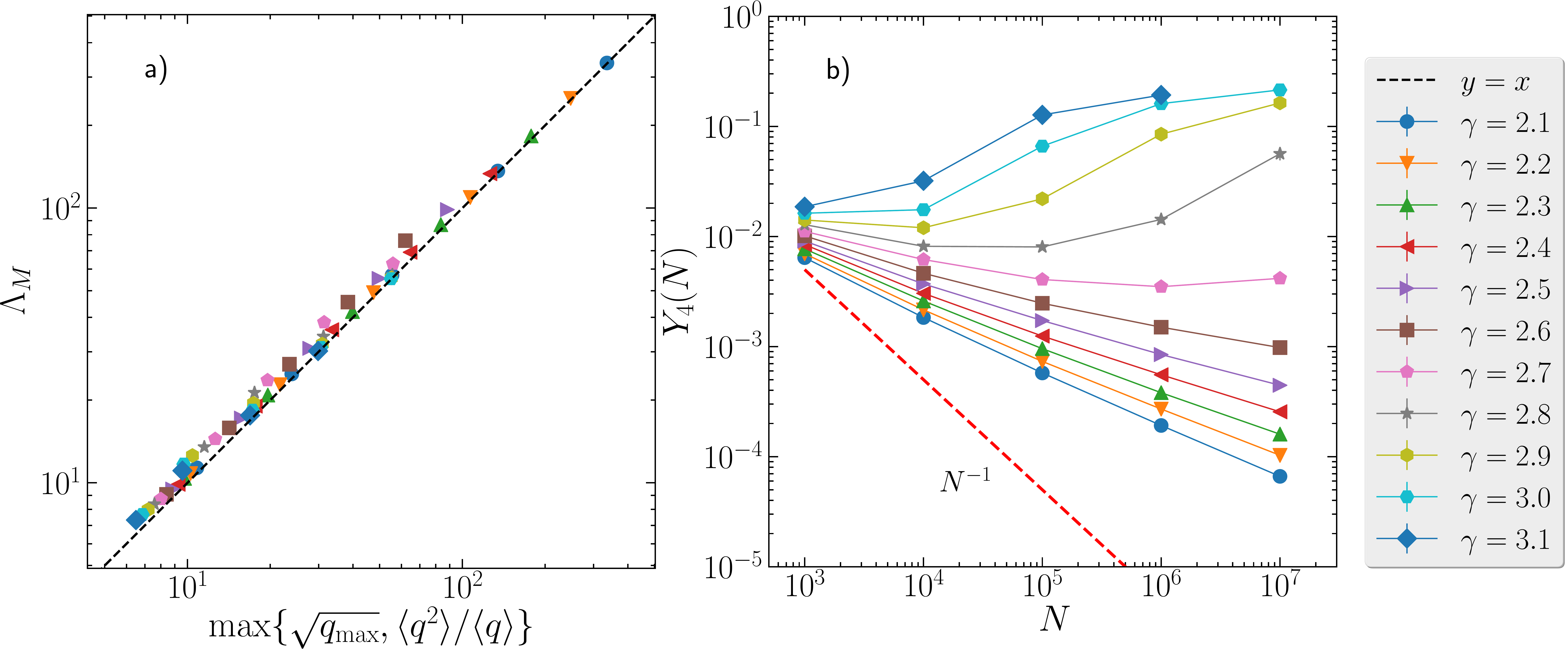}
  \caption{(\textbf{a}) Largest eigenvalue $\Lambda_M$ as a function of
    $\max \{ \sqrt{\km}, \av{q^2}/\av{q}\}$ in uncorrelated power-law
    UCM networks with different degree exponent $\gamma$ and network
    size $N$. (\textbf{b}) Inverse participation ratio $Y_4(N)$ as a
    function of $N$ in uncorrelated power-law UCM networks with
    different degree exponent $\gamma$.  Each point in both graphs
    corresponds to an average over $100$ independent network
    realization. Error bars are smaller than symbol sizes. Networks have
    a minimum degree $m=3$.}
  \label{fig:checkingCLV}
\end{figure}

It is important to remark that while Eq.~(\ref{eq:1}) holds for
asymptotically large networks, its applicability to networks of finite
(yet huge) size should not be taken for granted.  The exact
Eq.~(\ref{Lambda_M}) does not provide predictions for a wide
(size-dependent) range of values of the ratio $\CLV$. As shown in
Appendix~\ref{CLVconditions}, uncorrelated power-law distributed
networks fall within this range for the span of network sizes usually
considered in computer simulations, so that Eq.~(\ref{Lambda_M}) does
not provide predictions about any uncorrelated power-law networks that
can be numerically simulated.  Nevertheless, Eq.~(\ref{eq:1}), which is
a nonrigorous generalization of the exact Eq.~(\ref{Lambda_M}), turns
out to be very accurate for random uncorrelated static networks even of
small size.  Indeed, in Fig.~\ref{fig:checkingCLV}a we present a scatter
plot of $\Lambda_M$ computed using the power iteration
method~\cite{golub2012matrix} in random uncorrelated power-law networks
generated with the uncorrelated configuration model
(UCM)~\cite{Catanzaro05}, for different values of the degree exponent
$\gamma$ and network size $N$, as a function of the numerically
estimated value of $\max \{ \sqrt{\km}, \av{q^2}/\av{q}\}$. The
agreement with Eq.~(\ref{eq:1}) (in the following denoted as CLV theory)
is almost perfect, with only very small deviations for the smallest
network sizes.

\begin{figure}[t]
  \centering
  \includegraphics[width=0.7\textwidth]{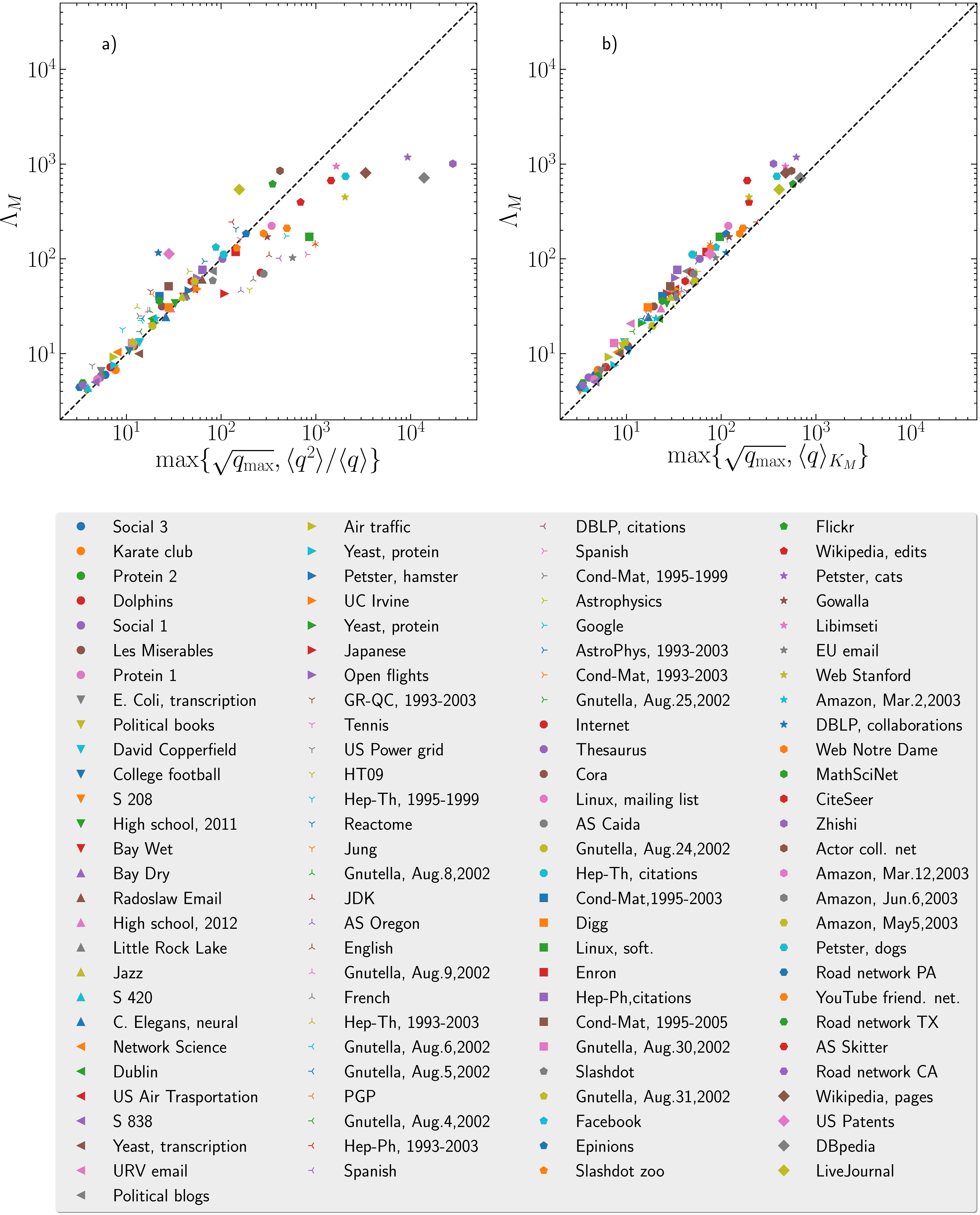}
  \caption{(\textbf{a}) Largest eigenvalue $\Lambda_M$ as a function of
    $\max \{ \sqrt{\km}, \av{q^2}/\av{q}\}$ computed for 109 different
    real-world networks.  (\textbf{b}) Largest eigenvalue $\Lambda_M$ as
    a function of $\max \{ \sqrt{\km}, \avK \}$ computed for the same
    real-world networks. Networks are ordered by increasing network
    size.}
  \label{fig:checkingLemma}
\end{figure}

In order to test the generality of CLV theory beyond uncorrelated
networks, we have also considered a large set of 109 real-world networks
(the same considered in Ref.~\cite{Radicchi15}, see this reference for
network details), of widely different origin, size and topological
features.  In Fig~\ref{fig:checkingLemma}a we plot the LEV of these
networks as a function of the numerically estimated quantity
$\max \{ \sqrt{\km}, \av{q^2}/\av{q}\}$. The result is quite clear:
While in some cases the CLV prediction works well, in others it provides
a overestimation of the actual value of the LEV that can be larger than
one order of magnitude.  This discrepancy is particularly strong in the
case of the \texttt{Zhishi}, \texttt{DBpedia}, and \texttt{Petster,
  cats} networks, but it is considerable in a large number of other
cases.

\section{Generalized formula for the largest eigenvalue}
\label{sec:gener-scal-larg}

We can understand the origin of the violations of the CLV formula
observed in Fig.~\ref{fig:checkingLemma}a and provide an improved
version, by reconsidering the observations made in
Refs.~\cite{Castellano2012,Pastor-Satorras2016}.  In these works it is
shown that the two types of scaling of the LEV in the CLV formula for
uncorrelated networks (either proportional to $\AV$ or to $\sqrt{\km}$)
are the manifestation of the two alternative ways in which the PEV can
become localized in the network (see
Appendix~\ref{sec:eigenv-local-inverse}).  For $\gamma>5/2$ the PEV is
localized around the node with largest degree in the network (the hub)
and the scaling of $\Lambda_M$ is given by $\sqrt{\km}$; for
$\gamma < 5/2$, on the other hand, the PEV becomes localized (in the
sense discussed in Ref.~\cite{Pastor-Satorras2016}) on the core of nodes
of maximum index $K_M$ in the $K$-core decomposition of the
network~\cite{Seidman1983269,Dorogovtsev2006} (see
Appendix~\ref{sec:k-core-decomposition}); the associated LEV is then
given by $\av{q^2}/\av{q}$. This picture is confirmed in
Fig.~\ref{fig:checkingCLV}b, where we study the localization of the PEV,
of components $\{ f_i \}$ assumed to be normalized as
$\sum_i f_i^2 = 1$, in random uncorrelated networks. The analysis is
performed by plotting the inverse participation
ratio~\cite{Goltsev12,2014arXiv1401.5093M,Pastor-Satorras2016} $Y_4(N)$
as a function of the network size (see
Appendix~\ref{sec:eigenv-local-inverse}). As we can check, for
$\gamma > 5/2$, $Y_4(N)$ goes to a constant for $N\to\infty$, indicating
localization in a finite set of nodes, that coincide with the hub and
its immediate neighbors. On the other hand, for $\gamma < 5/2$ the
inverse participation ratio decreases algebraically with network size,
with an exponent $\alpha$ smaller than $1$, indicating localization in a
sub-extensive set of nodes, which coincide with the maximum
$K$-core~\cite{Pastor-Satorras2016}. Additional evidence is presented in
Appendix~\ref{PEVweights}.

This observation can be interpreted in the following terms: The actual
value of the LEV in the whole network is the result of the competition
among two different subgraphs.  The node with largest degree $\km$ (the
hub), together with its immediate neighbors, form a star graph which, in
isolation, has a largest eigenvalue given by
$\Lambda_M^{(h)}=\sqrt{\km}$.  On the other hand, the maximum $K$-core,
of index $K_M$, is a densely interconnected, essentially
degree-homogeneous subgraph~\cite{Castellano2012}.  As such, its largest
eigenvalue is given by its internal average degree,
$\Lambda_M^{(K_M)} \approx \av{q}_{K_M}$. In the case of uncorrelated networks,
this average degree is well approximated by
$\AV$~\cite{Dorogovtsev2006}.  These two subgraphs, and their respective
largest eigenvalues, $\Lambda_M^{(h)}$ and $\Lambda_M^{(K_M)}$, compete
in order to set the scaling of the LEV of the whole network: The global
LEV coincides with the subgraph LEV that is larger.

We hypothesize that for generic networks, and also for correlated ones,
the same competition sets the overall LEV value.  The largest eigenvalue
of the star graph centered around the hub is trivially still equal to
$\sqrt{\km}$. What changes in general topologies, and in particular in
correlated networks, is the expression of the largest eigenvalue
associated to the maximum $K$-core.  One can realistically assume that
the maximum $K$-core is in general degree-homogeneous (see the
heterogeneity parameter of the maximum $K$-core or real-world networks
in Table~\ref{tab:myfirstlongtable}, which is, except in one case,
smaller than $1$).  What cannot be taken for granted in general is the
identification between $\avK$ and $\AV$. We thus conjecture that the LEV
in generic networks can be expressed as
\begin{equation}
  \Lambda_M \approx \max \{\sqrt{\km},\av{q}_{K_M}\}.
  \label{conjecture}
\end{equation}
Eq.~(\ref{conjecture}) is the central result of our paper.
Note Eq.~(\ref{conjecture}) is valid in full generality for any
network if the approximation sign $\approx$ is replaced by $\ge$,
as Rayleigh's inequality guarantees that the largest eigenvalue of
any subgraph is a lower bound for the whole network~\cite{PVM_graphspectra}.
Our conjecture here is that this lower bound is also
a very good approximation, in the sense that $\Lambda_M$ differs
from the lower bound by a factor of the order of few units.

Note that Eq.~(\ref{conjecture}) includes Eq.~(\ref{eq:1}) as a
particular case when $\av{q}_{K_M} \approx \AV$, which is true in
uncorrelated networks~\cite{Dorogovtsev2006}.  Moreover,
Eq.~(\ref{conjecture}) is in agreement with some known exact results for
specific classes of networks.  A simple example is provided by random
$q$-regular-graphs.  They have LEV equal to $q$, which is also the
average degree of the max $K$-core, and is larger than
$\sqrt{q_{max}}=\sqrt{q}$.  A less trivial example is given by random
trees grown according to the linear preferential attachment rule (see
Section~\ref{sec:case-line-pref}). Bhamidi et al.~\cite{Bhamidi2012}
show that in this case the LEV is exactly $\sqrt{\km}$, the value
predicted by Eq.~(\ref{conjecture}) since by construction
$\av{q}_{K_M}=2$.  Another related exact result concerns the $G(N,p)$
(Erd\"os-R\'enyi) random network, for which Krivelevich and
Sudakov~\cite{Krivelevich03} have proven that
$\Lambda_M = (1+o(1))\max\{\sqrt{q_{\max}}, Np \}$ where the term $o(1)$
tends to zero as the $\max\{\sqrt{q_{\max}}, Np \}$ tends to infinity
and $Np$ is the average degree $\av{q}$.  According to
Ref.~\cite{Dorogovtsev2006}, for Erd\"os-R\'enyi networks, the highest
$K$-core is linear with the average degree, $K_M \sim 0.78 \av{q}$, and
the mean degree of the $K$-core depends weakly on the core and
$\av{q}_K \simeq \av{q}$.  Hence our Eq.~(\ref{conjecture}) agrees with
the result of Krivelevich and Sudakov for Erd\"os-R\'enyi networks.

In Fig.~\ref{fig:checkingLemma}b we check the validity of the proposed
generalized scaling for the LEV in the case of the 109 real-world
networks considered above.  Comparing with
Fig.~\ref{fig:checkingLemma}a, the generalized formula provides a much
better overall fitting to the real value of the LEV than the original
CLV expression, and therefore represents a better prediction for the
behavior of this quantity. The overall improvement of our prediction
versus the original CLV one can be established by comparing the absolute
relative errors, with respect to actual measured LEVs, of the values
predicted by Eq.~(\ref{eq:1}) and Eq.~(\ref{conjecture}), respectively.
The average relative error for Eq.~(\ref{eq:1}) is $1.213$, with
standard deviation $3.285$, and a maximum of $26.686$; for
Eq.~(\ref{conjecture}), the average is $0.282$, with standard deviation
$0.154$ and maximum $0.719$.  We conclude that Eq.~(\ref{conjecture})
provides an excellent approximation for the LEV value of an extremely
broad variety of networks. Additional evidence of its predictive power
is presented in the next Section.

Despite this vast generality, there are however particular classes of
networks for which the lower bound is not tight and
Eq.~(\ref{conjecture}) is not a good approximation. These cases are
examined in our discussion, Section~\ref{sec:Discussion}.

\section{The case of linear preferential attachment networks}
\label{sec:case-line-pref}

Growing network models provide a particularly interesting testbed for
the conjecture presented above.  We focus in particular on linear
preferential attachment (LPA)
networks~\cite{Barabasi:1999,Dorogovtsev2000}, generated starting from a
fully connected nucleus of $m+1$ nodes and adding at every time step a
new node with $m$ new edges connected to $m$ old nodes. For the vertex
introduced at time $t$, each of its emanating edges is connected to an
existing vertex $s$, introduced at time $s<t$, with probability
\begin{equation}
  \Pi_s(t) = \frac{q_s(t) + a}{\sum_j[q_j(t) + a]},
\end{equation}
where $q_s(t)$ is the degree, measured at time $t$, of the node
introduced at time $s$.  The factor $a$ takes into account the possible
initial attractiveness of each node, prior to receiving any connection.
Large LPA networks are characterized by a power-law degree distribution
$P(k) \sim k^{-\gamma}$, with a degree exponent
$\gamma = 3 + \frac{a}{m}$~\cite{Barrat2005} and average degree
$\av{q} = 2m$.  It is thus possible to tune the degree exponent in the
range $2 < \gamma < \infty$ by changing the attractiveness parameter in
the range $-m < a < \infty$.  The power-law form extends up to the
maximum degree $\km$ that depends on $N$ as $\km \sim N^{1/(\gamma-1)}$
for all values of $\gamma$. LPA networks are further characterized by
the presence of degree correlations~\cite{assortative}: The average
degree of the nearest neighbors of nodes of degree $q$,
$\overline{q}_{nn}(q)$~\cite{alexei} is of the form
$\overline{q}_{nn}(q) \sim q^{-3 + \gamma}$ for $\gamma <3$, and
$\overline{q}_{nn}(q) \sim \ln q$ for $\gamma >
3$~\cite{Barrat2005}. See Appendix~\ref{sec:build-line-pref} for a
practical implementation of this model.

By their very construction LPA networks lack a nontrivial $K$-core structure,
since the iterative procedure to determine $K$-shells for $K>m$ removes all
nodes by exactly reversing the growth process.  Therefore, in LPA
networks, all nodes belong to the same $K=m$ shell, where $m$ is the
minimum degree in the network. We thus have
$\avK = \av{q} = 2m \ll \sqrt{\km}$ even for modest values of $N$, and
according to our generalized prediction, the LEV should be approximately
$\sqrt{\km}$ for all values of $\gamma$, in stark opposition to the
original CLV formula, that still predicts in Eq.~(\ref{eq:4})
different expressions for $\gamma<5/2$ and $\gamma > 5/2$. This scaling
$\Lambda_M \sim \sqrt{\km}$ has been exactly demonstrated for the case
$\gamma=3$, corresponding to the so-called Barabasi-Albert
model~\cite{Barabasi:1999} in Ref.~\cite{Flaxman2005}. Here we extend
this form for all values of $\gamma$ in LPA networks.

\begin{figure}[t]
  \centering
  \includegraphics[width=0.8\textwidth]{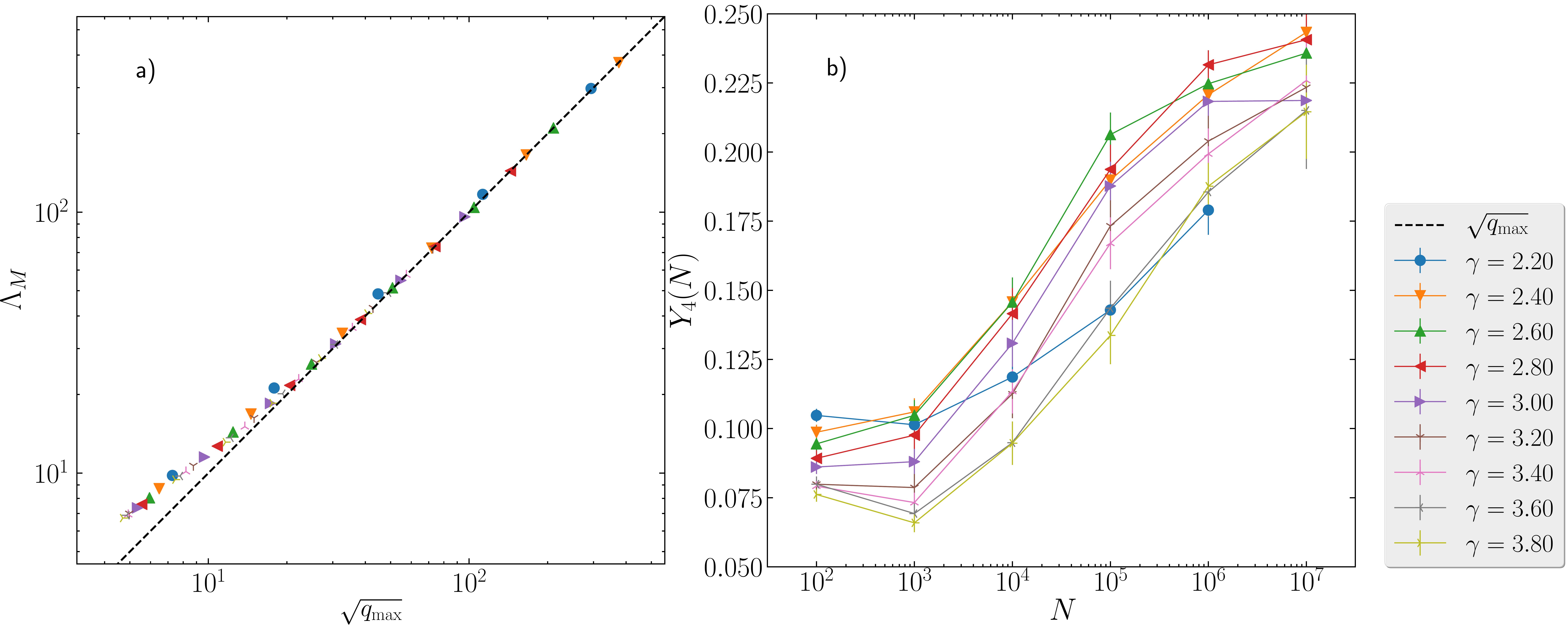}
  \caption{(\textbf{a}) Largest eigenvalue $\Lambda_M$ as a
    function of $\sqrt{\km}$ in LPA networks with different degree
    exponent $\gamma$. (\textbf{b}) Inverse participation ratio $Y_4(N)$
    as a function of $N$ in LPA networks with different degree exponent
    $\gamma$. Each point in both graphs corresponds to an average over
    100 independent network realization. Error bars are smaller than
    symbol sizes.}
  \label{fig:checkingLemmaLPA}
\end{figure}

\begin{figure}[t]
  \centering
  \includegraphics[width=0.8\textwidth]{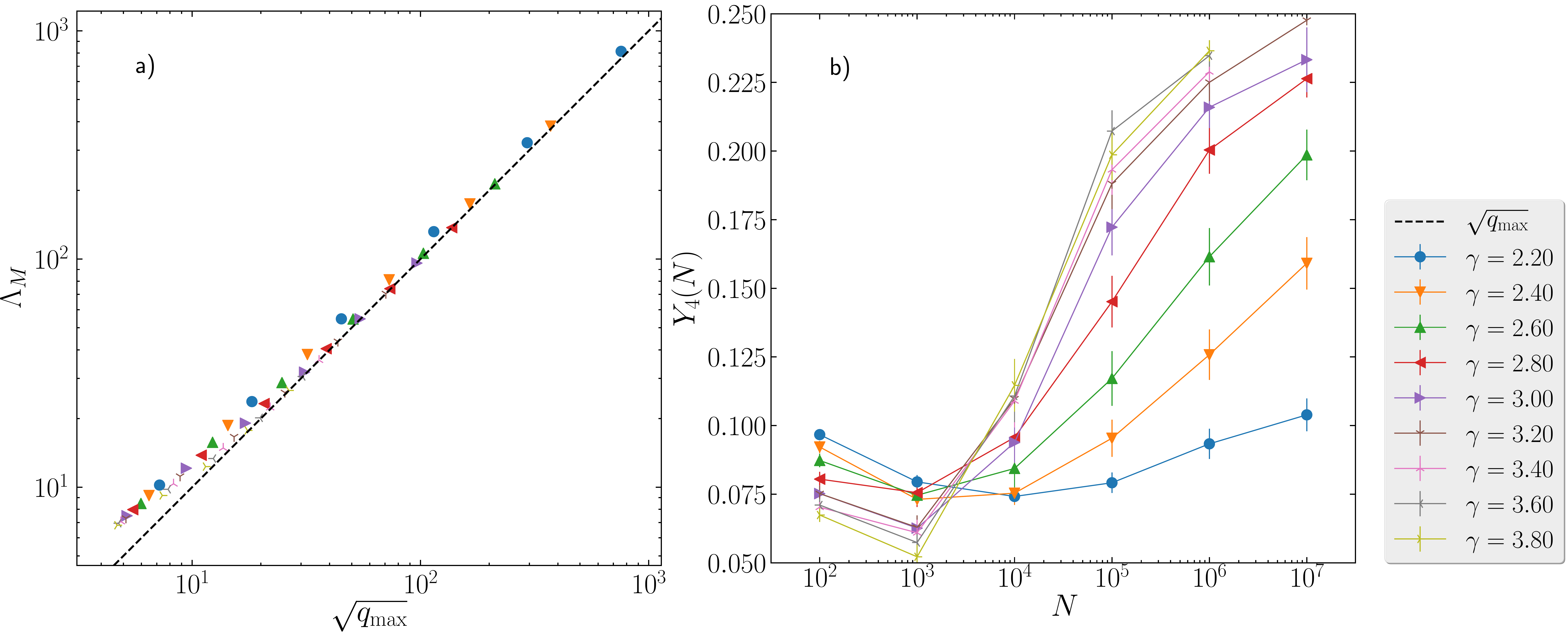}
  \caption{(\textbf{a}) Largest eigenvalue $\Lambda_M$
    as a function of $\sqrt{\km}$ in rewired LPA networks with different
    degree exponent $\gamma$. (\textbf{b}) Inverse participation ratio
    $Y_4(N)$ as a function of $N$ in rewired LPA networks with different
    degree exponent $\gamma$. Each point in both graphs corresponds to
    an average over 100 independent network realization. Error bars are
    smaller than symbol sizes.}
  \label{fig:checkingLemmaLPArewired}
\end{figure}

In Fig.~\ref{fig:checkingLemmaLPA}a we plot the largest eigenvalue
obtained in LPA networks with different size and degree exponent
$\gamma$, as a function of $\sqrt{\km}$. As we can observe, after a
short preasymptotic regime for small network sizes (small $\km$),
$\Lambda_M$ grows as $\sqrt{\km}$ for all values of $\gamma$,
independently of the factor $\av{q^2}/\av{q}$. Interestingly, for all
values of $\gamma$, the LEV falls onto the same universal curve
asymptotically approaching $\sqrt{\km}$, which indicates that this
functional form is moreover independent of degree correlations, which
change continuously with $\gamma$ in LPA networks~\cite{Barrat2005}. We
conclude that, in perfect agreement with our conjecture, the spectral
properties of LPA networks are ruled only by the hub. This implies
additionally that the PEV is localized around the hub.  This fact is
verified in Fig.~\ref{fig:checkingLemmaLPA}b, where we plot the inverse
participation ratio $Y_4(N)$ as a function of $N$.  In
Fig.~\ref{fig:checkingLemmaLPA}b it turns out clearly that $Y_4(N)$ goes
to a constant for $N\to\infty$, for any degree exponent $\gamma$ and for
sufficiently large $N$, indicating that PEV always becomes localized on
a set of nodes of finite size (not increasing with $N$): The hub and its
immediate neighbors.  Further evidence about the localization is
provided in Appendix~\ref{sec:princ-eigenv-local}.

In LPA networks, the lack of a $K$-core structure is a fragile property,
since reshuffling connections while preserving the degree of each 
node~\cite{maslov02} may induce some $K$-core structure. 
This emerging $K$-core
structure is however not able to restore the scaling predicted by
Eq.~(\ref{eq:1}) (see Appendix~\ref{sec:k-core-structure}).  As
Fig.~\ref{fig:checkingLemmaLPArewired}a shows, reshuffling does not
alter the overall behavior, apart from minimal changes: the LEV still
scales asymptotically as $\sqrt{\km}$ for any $\gamma$, while the PEV is
still asymptotically localized around the hub, as the inverse
participation ratio tending to a constant for large network sizes shows,
Fig.~\ref{fig:checkingLemmaLPArewired}b.

\section{Consequences for dynamics on networks}
\label{sec:cons-dynam-netw}

The spectral properties of the adjacency matrix determine the behavior
of many dynamical processes mediated by topologically complex contact
patterns~\cite{Castellano2010,Restrepo2005,Restrepo2008,Pomerance2009,Larremore2011}.
Here we show the consequences that the topological properties uncovered
above have for two highly relevant types of dynamics.

\subsection{Epidemic spreading}

The Susceptible-Infected-Susceptible (SIS) model is one of the simplest
and most fundamental models for epidemic spreading~\cite{anderson92}
(see Appendix~\ref{sec:susc-infect-susc} for details), showing an
epidemic threshold $\lambda_c$ separating a regime where epidemics get
rapidly extinct from a regime where they affect a finite fraction of the
system. The dependence of this threshold on the network topology is well
approximated by the so-called Quenched Mean-Field theory (QMF) (see
Appendix~\ref{sec:susc-infect-susc}), predicting it to be equal to the
inverse of the LEV
\begin{equation}
  \lambda_c=\frac{1}{\Lambda_M}.
  \label{lambda_c}
\end{equation}
Inserting into this expression the LEV scaling form given by
Eq.~(\ref{eq:1}) in the case of random uncorrelated static networks, we
see that the threshold always vanishes on power-law distributed networks
in the thermodynamic limit, with different scalings depending on the
value of $\gamma$~\cite{Castellano2010}. For $\gamma<5/2$ the expression
coincides with the one predicted by the Heterogeneous Mean-Field (HMF)
theory~\cite{pv01a} (see Appendix~\ref{sec:susc-infect-susc}), while HMF
theory is violated for $\gamma>5/2$.  In LPA networks $\Lambda_M$ is,
for any $\gamma$, given by $\sqrt{\km}$, so that Eq.~(\ref{lambda_c})
predicts a vanishing of the epidemic threshold qualitatively different
from the one on uncorrelated networks for $\gamma<5/2$. In particular,
the approach to zero in the thermodynamic limit should be
\textit{slower} in LPA networks than in static uncorrelated networks
with the same $\gamma$.

\begin{figure}[t]
  \centering
  \includegraphics[width=0.8\textwidth]{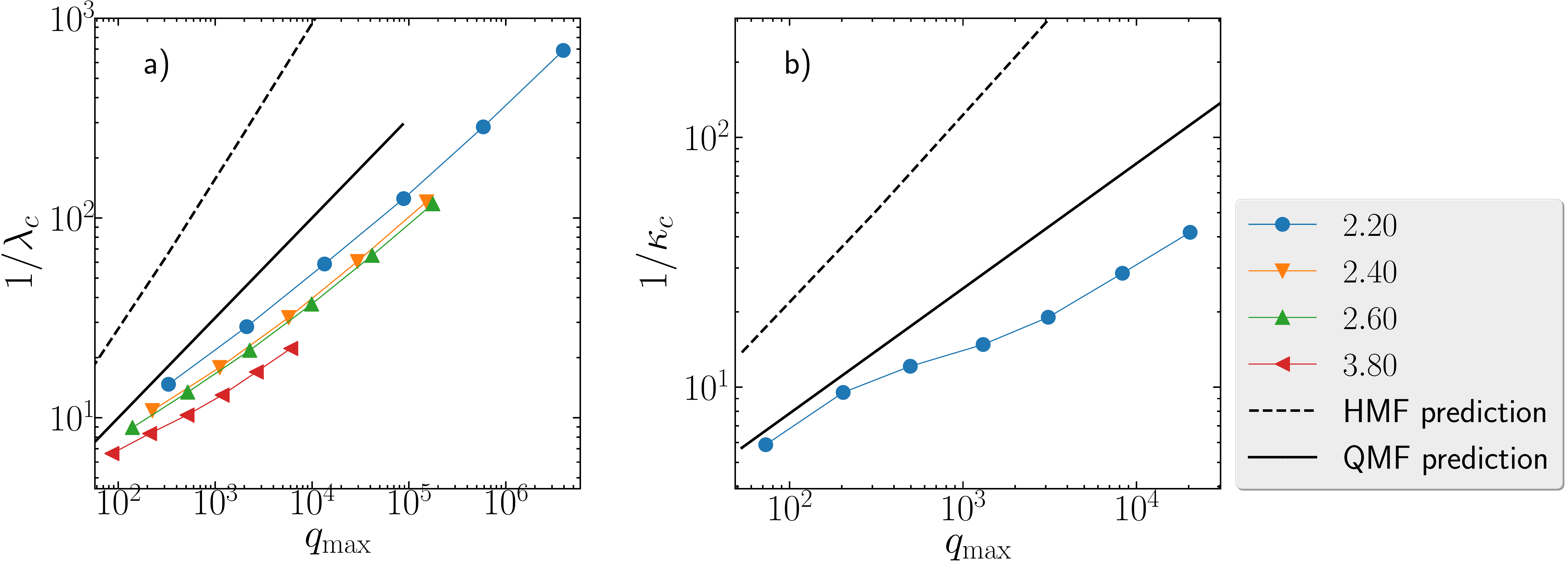}
  \caption{(\textbf{a}) Numerical estimate of the inverse epidemic
    threshold $1/\lambda_c$ in LPA as a function of $\km$, for various
    values of the exponent $\gamma$.  We consider networks of sizes from
    $N=10^2$ to $N=10^8$. (\textbf{b}) Numerical estimate of the
    synchronization threshold $\kappa_c$ for $\gamma=2.2$ as a function
    of $q_{max}$.  System size ranges from $N=300$ to $N=300000$. In
    both plots the dependence for $\gamma=2.2$ predicted by the QMF
    theory is represented by a thick dashed straight line, and the HMF
    prediction is represented by a dash-dotted line}
  \label{fig:lifetime}
\end{figure}

In order to check this picture, we perform numerical simulations of the
SIS model on LPA networks of different degree exponent $\gamma$, and
determine the threshold using the lifespan method (see
Appendix~\ref{sec:susc-infect-susc}).  In Fig.~\ref{fig:lifetime}a we
plot the numerically estimated threshold as a function of $\km$.  We
find that the theoretical expectation is followed only approximately:
the slopes are smaller than 1 in all cases, the more so for larger
values of $\gamma$. However, this discrepancy is a finite size effect:
as the system size grows the effective slope grows. Asymptotically for
large $N$ the threshold always vanishes as $\sqrt{\km}$, at variance
with what happens for uncorrelated static networks for $\gamma<5/2$. The
comparison with the slope predicted by HMF theory for $\gamma=2.2$
(dashed line) clearly shows the failure of the latter.  Hence the
remarkable conclusion that on LPA networks the epidemic threshold
vanishes asymptotically for any $\gamma$, but it never vanishes as
predicted by HMF theory, at odds with what happens on static
uncorrelated networks.

\begin{figure}[t]
  \centering
  \includegraphics[width=0.7\textwidth]{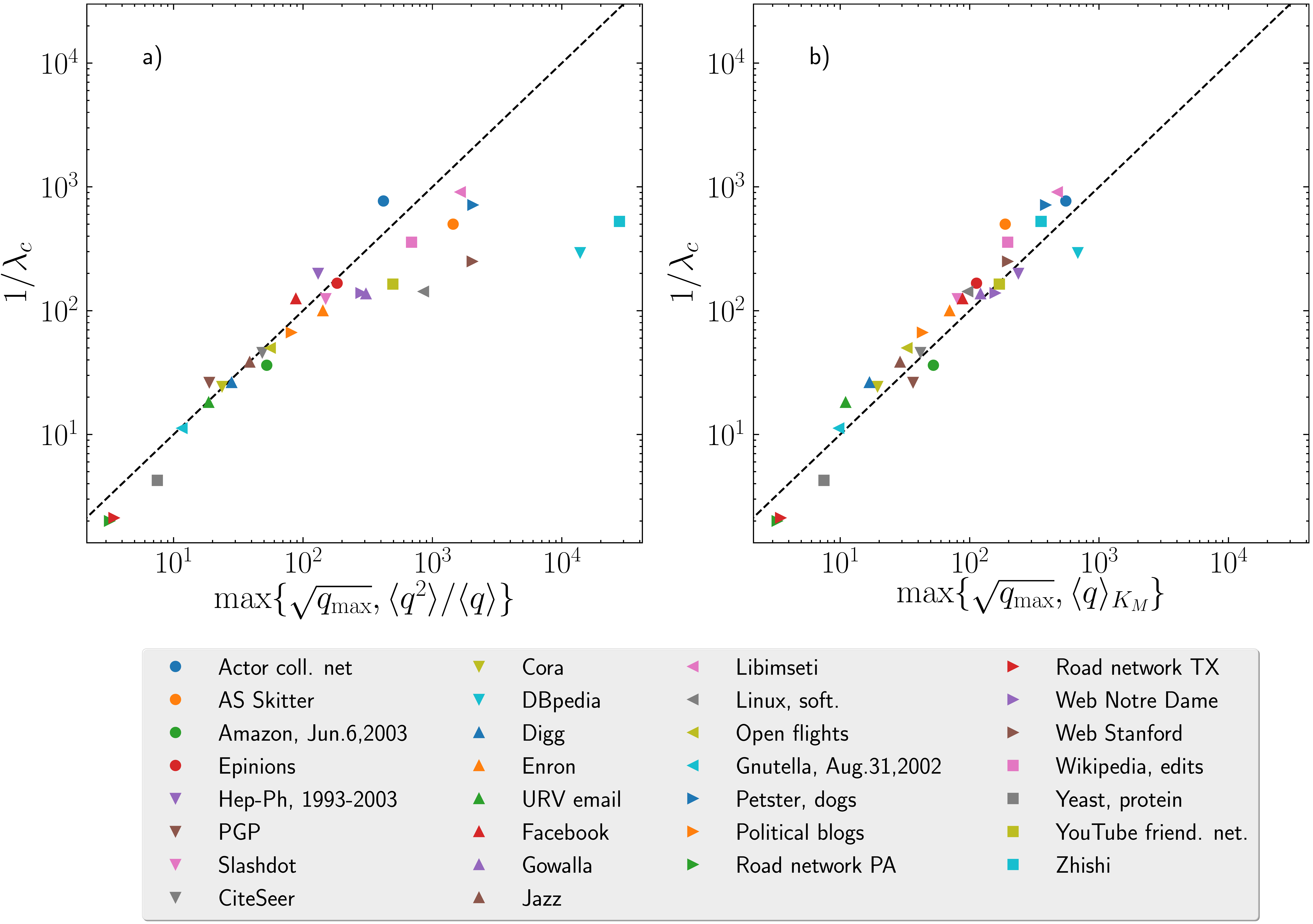}
  \caption{(\textbf{a}) Numerical estimate of the inverse epidemic
    threshold $1/\lambda_c$, in a subset of the real-world networks
    considered, as a function of the inverse largest eigenvalue
    approximation $\max\{ \sqrt{\km}, \AV \}$.  (\textbf{b}) Numerical
    estimate of the inverse epidemic threshold $1/\lambda_c$ in
    real-world networks as a function of the inverse improved largest
    eigenvalue approximation $\max\{ \sqrt{\km}, \avK \}$. Data for the
    networks \texttt{Zhishi}, \texttt{Web Notre Dame}, \texttt{Road
      network TX} and \texttt{Road network PA} are lower bounds to the
    real threshold, due to computing time limitations.}
  \label{fig:lifetimeRealnets}
\end{figure}

In the case of real-world networks, our proposed estimate for the
scaling of the largest eigenvalue again provides a much better overall
prediction for the threshold in the SIS model, see
Fig.~\ref{fig:lifetimeRealnets}, where we compare it with the original
CLV prediction. As we can see, in cases where the CLV prediction is off
by orders of magnitude, our improved scaling form leads to a much better
threshold prediction. As an estimate of the overall goodness of the
prediction, the mean relative error for the CLV predictions is
$4.20$ (standard deviation $11.92$, maximum $52.02$).
while the predictions of Eq.~(\ref{conjecture}) give much smaller values
(mean $0.37$, standard deviation $0.24$, maximum $1.33$).

\subsection{Synchronization}

Kuramoto dynamics~\cite{RevModPhys.77.137} (see
Appendix~\ref{sec:kuram-synchr-dynam}) is the paradigmatic model for the
study of synchronization among weakly coupled oscillators, with
applications ranging from neural networks to charge-density waves. Its
behavior in networks has been investigated in great
detail~\cite{Arenas2008,Rodrigues2016}, showing the existence of a
synchronization threshold for a coupling parameter, $\kappa_c$,
separating a random from a synchronized phase.  Concerning the
synchronization threshold, the standard approach is the one in
Ref.~\cite{Restrepo2005}, predicting a synchronized state to appear when
the coupling $\kappa$ among oscillators is larger than the critical
value
\begin{equation}
  \kappa_c = \frac{k_0}{\Lambda_M},
\end{equation}
where $k_0=2/[\pi g(0)]$ and $g(\omega)$ is the frequency distribution
of individual oscillators (see Appendix~\ref{sec:kuram-synchr-dynam}).
To assess whether the generalized scaling just uncovered for $\Lambda_M$
on LPA networks has effects also for these dynamics, we perform
simulations of the Kuramoto model on growing networks and determine the
critical coupling $\kappa_c$ (see Appendix~\ref{sec:kuram-synchr-dynam}
for details).

Fig.~\ref{fig:lifetime}b clearly shows that also for these dynamics the
prediction given by the inverse of the LEV is qualitatively correct and,
as a consequence, for $\gamma<5/2$ the threshold vanishes more slowly
than what is predicted for random uncorrelated networks. We conclude
that also in this case the nature of the growing network, and in
particular the lack of a $K$-core structure, has profound consequences
for the dynamics mediated by the contact network.

\section{Discussion}
\label{sec:Discussion}

The generalized CLV conjecture we have exposed allows to fully clarify
the physical origin of the properties of the adjacency matrix largest
eigenpair in complex networks.  There are two subgraphs which determine
the LEV and the PEV in a large complex network: the hub with its spokes
and the densely mutually interconnected set of nodes singled out as the
maximum $K$-core~\footnote{Notice that in many cases the hub actually is
  part of the maximum $K$-core. Nevertheless there is a clear
  distinction between the case the PEV is localized on the hub and its
  immediate neighbors or the PEV is localized around the maximum
  $K$-core as a whole}.  Each of these two subgraphs has (in isolation)
an associated LEV: the hub is the center of a star graph and therefore
$\Lambda_M^{(h)}=\sqrt{\km}$; the maximum $K$-core is a homogeneous
graph and therefore $\Lambda_M^{(K_M)}\simeq\av{q}_{K_M}$.  The LEV of
the global topology is simply given by the largest of the two.  In
uncorrelated static networks the growth with $N$ of the two individual
LEVs depends on $\gamma$ and this gives rise to the change of behavior
occurring for $\gamma=5/2$, Eq.~(\ref{eq:4}).  In growing LPA networks
the $K$-core structure is by construction absent: The spectral
properties are dictated only by the hub (and this remains true also
after reshuffling).  In networks of any origin (and any correlation
level), the relation between the average degree of the max $K$-core and
$\AV$ may break down.  However it is still true that the LEV value is
the largest between $\Lambda_M^{(h)}$ and $\Lambda_M^{(K_M)}$.

This conjecture is clearly not a proof.  However, the understanding of
its conceptual origin allows to predict that it should hold for
practically all real-world networks.  Various different mechanisms 
may lead to its breakdown.  There could be an {\it inhomogeneous} 
max $K$-core in the network, so that $\Lambda_M^{(K_M)}$ is very different
from $\av{q}_{K_M}$.
There could be a third, different, type of subgraph, characterized 
by a LEV larger than both the others. 
Or the whole graph could have a LEV larger than the LEV of any
proper subgraph. An example of this last case is the complete bipartite 
network $K_{p,q}$: its LEV is$\sqrt{pq}$~\cite{PVM_graphspectra}, 
which can be much larger (assuming $p \le q$) than the value
$\max\{\sqrt{q}, p \}$ predicted by Eq.~(\ref{conjecture}).
All these mechanisms are in principle possible; however they
appear to be unlikely in real self-organized networks.

Our findings about spectral properties have immediate implications in
several contexts. We have shown that properties of dynamical processes
as general as epidemics and synchronization are deeply affected by which
subgraph determines the LEV. For example, another effect that can be
immediately predicted, is that removing the hub may completely disrupt
the dynamics when the LEV is given by $\Lambda_M^{(h)}$ while being
practically inconsequential in the other case.  Similar consequences are
expected to occur in general~\cite{Restrepo2008,Pomerance2009,
  Larremore2011,Kinouchi:2006aa}.  Another context where these results
may have implications is for centrality measures, many of which are
variations of the eigenvalue centrality~\cite{Bonacich72,Katz1953}.
Finally, it is worth to remark that the example of linear preferential
attachment networks clearly points out that the way a network is built
may have deep and unexpected implications for its structure and its
functionality.

\begin{acknowledgments}
   We acknowledge financial support from the Spanish MINECO, under
  projects FIS2013-47282-C2- 2 and FIS2016-76830-C2-1-P.
  R.P.-S. acknowledges additional financial support from ICREA Academia,
  funded by the Generalitat de Catalunya.
\end{acknowledgments}

\appendix

\section{Applicability of CLV exact results to finite networks}
\label{CLVconditions}

\begin{figure}[t]
  \centering
  \includegraphics[width=0.7\textwidth]{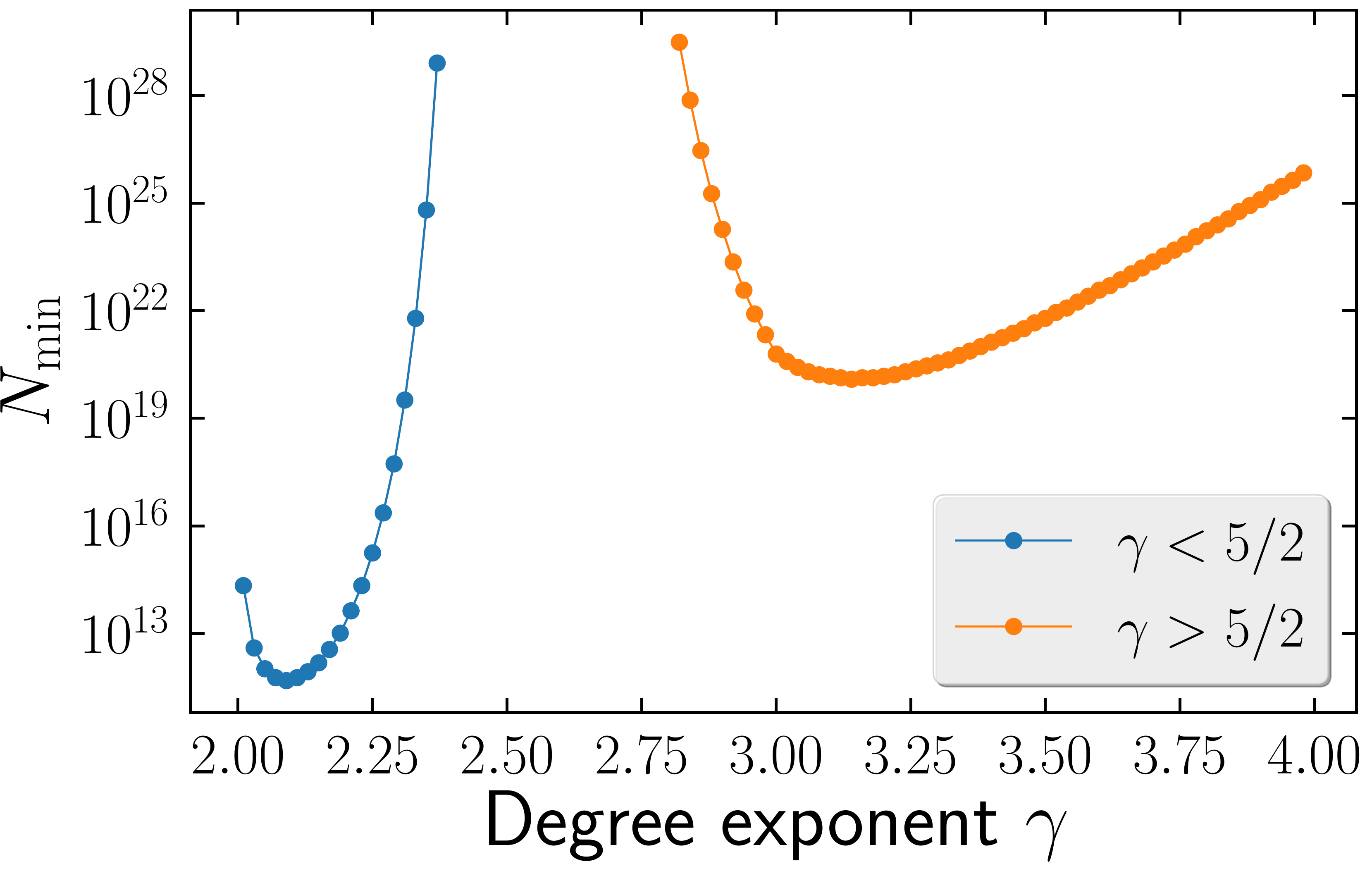}
  \caption{Minimum sizes $N_\mathrm{min}$ for the validity of
    Eq.~(\ref{Lambda_M}) in uncorrelated power-law networks as a
    function of the degree exponent $\gamma$. Values in the vicinity of
    $\gamma = 5/2$ not plotted are all larger than $10^{30}$.}
  \label{fig:lowerN}
\end{figure}

The exact result proved by Chung, Lu and Vu in Ref.~\cite{Chung03},
namely Eq.~(\ref{Lambda_M}), can be rewritten as
\begin{equation}
  \Lambda_M =
  \left \{ 
    \begin{array}{lcr}
      a_1 \sqrt{\km} &\;\mathrm{if}\;\;&  \frac{\av{q^2}}{\av{q}
                                         \sqrt{\km}} < \frac{1}{\ln^2(N)} \\ 
      a_2 \frac{\av{q^2}}{\av{q}} &\;\mathrm{if}\;\;&  \frac{\av{q^2}}{\av{q}
                                         \sqrt{\km}} > \ln(N) 
    \end{array}
  \right. .
  \label{Lambda_M_other}
\end{equation}
It therefore provides a prediction for the value of the LEV if the ratio
\begin{equation}
  \label{eq:6}
  \zeta(\gamma, N) \equiv \frac{\av{q^2}}{\av{q} \sqrt{\km}}
\end{equation}
is larger than $\ln(N)$ or smaller than $1/\ln^2(N)$. For very large
systems both $\AV$ and $\sqrt{\km}$ diverge and, if they scale with $N$
with different exponents (i.e. $\gamma \ne 5/2$), the logarithmic
factors are not asymptotically relevant: either the first or the second
of the conditions in Eq.~(\ref{Lambda_M}) is fulfilled.  However, for
finite values of $N$ there is a sizeable interval such that
Eq.~(\ref{Lambda_M}) does not strictly apply. In the case of uncorrelated
power-law networks with distribution
$P(q) = (\gamma-1)m^{\gamma-1} q^{-\gamma}$, we have, in the continuous
degree approximation
\begin{equation}
  \label{eq:3}
  \zeta(\gamma, N)  =
  \frac{\gamma-2}{3-\gamma}\frac{m}{\sqrt{\km}}
  \left[ \left( \frac{\km}{m}\right)^{3-\gamma} - 1\right],
\end{equation}
where $\km = N^{1/2}$ for $\gamma < 3$ and $\km = N^{1/(\gamma-1)}$ for
$\gamma>3$, $m$ is the minimum degree, which we take $m=3$, and in the
evaluation of $\av{q^2}$ we have taken the maximum degree $\km$ into
account.  Evaluating numerically $\zeta(\gamma, N)$ we can compute for
every value of $\gamma$ the minimum value of $N$ for the exact
expression Eq.~(\ref{Lambda_M}) to apply.  For $\gamma<5/2$,
$\zeta(\gamma, N)$ diverges with $N$: the prediction
$\Lambda_M \approx \AV$ in Eq.~(\ref{Lambda_M}) applies for
$N>N_\mathrm{min}$ defined by
$\zeta(\gamma, N_\mathrm{min}) = \ln(N_\mathrm{min})$.  On the other
hand, for $\gamma>5/2$, $\zeta(\gamma, N)$ tends to zero as the system
size diverges. Hence the prediction $\Lambda_M \approx \sqrt{\km}$ in
Eq.~(\ref{Lambda_M}) applies for $N>N_\mathrm{min}$, defined in this
case by $\zeta(\gamma, N_\mathrm{min}) = 1/\ln^2(N_\mathrm{min})$.  In
Fig.~\ref{fig:lowerN} we plot $N_\mathrm{min}$ as a function of
$\gamma$.
From the figure it turns out that the exact theoretical prediction
Eq.~(\ref{Lambda_M}) holds only for extremely large size (at least
$N>10^{11}$, but the bound is much larger for almost all values of
$\gamma$).  Networks of such size cannot be simulated with current
computer resources.  An improved analysis in Ref.~\cite{Chung03}
replaces $\ln(N)$ by $\ln(N)^{1/2}$ and $\ln(N)^2$ by $\ln(N)^{3/2}$ in
Eq.~(\ref{Lambda_M}). A similar analysis as performed above indicates
that this corrected version holds for sizes at least $N > 3\times 10^7$,
which is very close to the limit allowed by present computation systems.

\begin{figure*}[t]
\centering
\includegraphics[width=\textwidth]{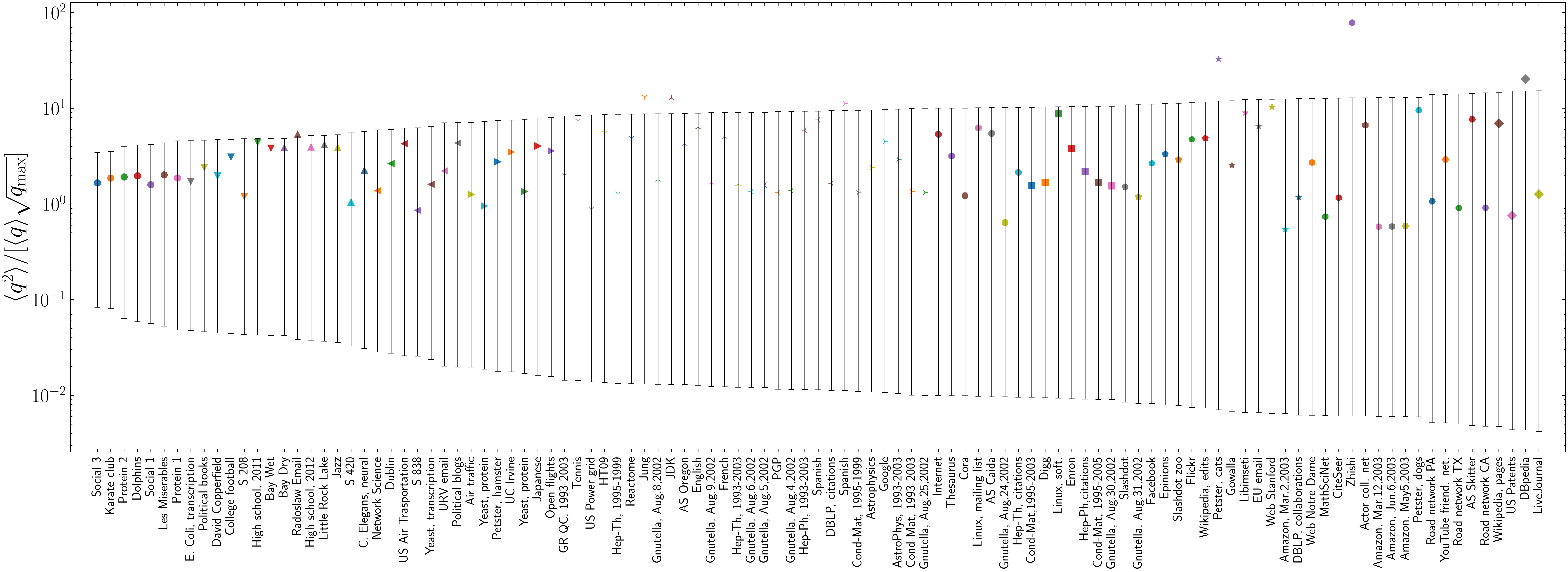}
\caption{For each of the real-world networks considered, we plot in
log scale a line indicating the interval where Eq.~(\ref{Lambda_M})
does not make any prediction. The symbols indicate the actual value
of $\CLV$ for each network.}
\label{Anti}
\end{figure*}

In the case of the real-world networks considered, we plot in
Fig.~\ref{Anti} along the $y$-axis a line between $\ln(N)$ and
$1/\ln^2(N)$, indicating the interval of values of $\CLV$ where
Eq.~(\ref{Lambda_M}) does not strictly apply.  The symbols indicate the
actual value of the ratio $\CLV$ in each network. It turns out that for
102 networks out of 109 the actual value falls in the inapplicability
interval. Hence for the vast majority of real networks the exact result
in Eq.~(\ref{Lambda_M}) does not, strictly speaking, allow to make any
prediction.

\section{Eigenvector localization and the inverse participation ratio}
\label{sec:eigenv-local-inverse}

The concept of the localization of the principal eigenvector $\{ f_i\}$
translates in determining whether the value of its normalized components
(satisfying $\sum_i f_i^2 = 1$) is evenly distributed among all nodes in
the network or it attains a large value on some subset of nodes $V$ and
is much smaller in all the rest.  In the first case $f_i \sim N^{-1/2}$,
$\forall i$ and the network is not localized.  In the second case
$f_i \sim N_V^{-1/2}$, for $i \in V$ and $f_i \sim 0$, for $i \notin V$,
were $N_V$ is the size of the localization subset $V$.

The presence of localization in the PEV can be easily assessed in
ensembles of networks of variable size $N$ by studying the inverse
participation ratio (IPR), defined
as~\cite{Goltsev12,2014arXiv1401.5093M}
\begin{equation}
  \label{eq:2}
  Y_4(N) = \sum_{i=1}^N [f_i]^4,
\end{equation}
as a function of $N$, and fitting its behavior to a power-law decay of
the form~\cite{Pastor-Satorras2016}
\begin{equation}
  \label{eq:5}
  Y_4(N) \sim N^{-\alpha}.
\end{equation}
If the PEV is delocalized, with $f_i \sim N^{-1/2}$, $\forall i$, the
exponent $\alpha$ is equal to $1$. Any exponent $\alpha<1$ indicates the
presence of some form of eigenvector localization, taking place in a
sub-extensive set of nodes, of size $N_V \sim N^{\alpha}$. In the
extreme case of localization on a single node, or a set of nodes with
fixed size, we have $\alpha=0$ and $Y_4(N) \sim \mathrm{const.}$.

\section{The $K$-core decomposition}
\label{sec:k-core-decomposition}

The $K$-core decomposition~\cite{Seidman1983269} is an iterative
procedure to classify vertices of a network in layers of increasing
density of mutual connections.  Starting with the whole graph one removes the
vertices with only one connection (degree $q=1$). This procedure is then
repeated until only nodes with degree $q \ge 2$ are left. The removed
nodes constitute the $K$=1-shell and those remaining are the
$K$=2-core. At the next step all vertices with degree $q=2$ are removed,
thus leaving the $K$=3-core.  The procedure is repeated iteratively. The
maximum $K$-core (of index $K_M$) is the set of vertices such that one
more iteration of the procedure removes all of them.  Notice that all
vertices of the $K$-core of index $K$ have degree larger than or equal
to $K$.

\section{Different localizations for UCM networks}
\label{PEVweights}
\begin{figure}[t]
  \centering
  \includegraphics[width=0.6\textwidth]{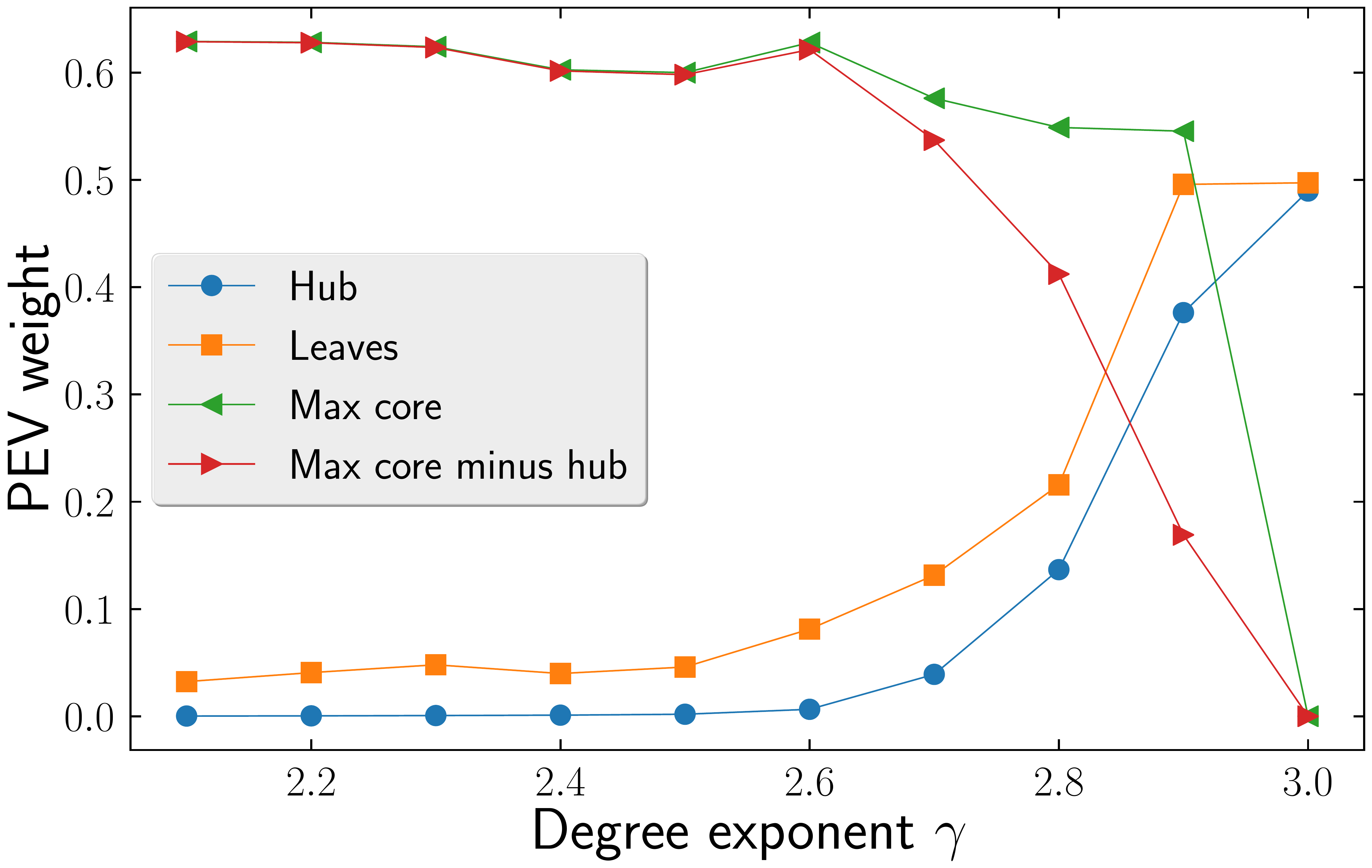}
  \caption{PEV weight concentrated on various subgraphs for UCM networks
    of size $N=10^7$ with different values of $\gamma$.}
  \label{weightsplot}
\end{figure}

For UCM networks the localization of the PEV in different subgraphs
depending on whether $\gamma$ is larger or smaller than 5/2 can be
exposed by plotting the weights concentrated on the subgraphs 
as a function of $\gamma$ (see Fig.~\ref{weightsplot}).
In this figure we have set the weight of the maximum $K$-core equal to
zero for $\gamma=3$, since by construction it coincides with the whole
network, and trivially contains all the weight for the PEV.

It is clear that for $\gamma<5/2$ the weight is concentrated on the
max $K$-core, while the hub plays no role. 
For $\gamma>5/2$ the opposite scenario applies:
the hub plus its nearest neighbors (leaves) bear 
most of the PEV weight, while the max $K$-core (or the max $K$-core
minus the hub, in case the latter belongs to the former) has vanishing
weight concentrated on it.
Strong finite size effects smoothen the change of behavior for 
$\gamma$ between $5/2$ and $3$; but changing the system size
(not shown) one can extrapolate that for asymptotically large systems
the picture is the one just described.

\section{Building linear preferential attachment networks}
\label{sec:build-line-pref}

Given the mapping of LPA networks with the Price model \cite{Price75},
LPA networks can be easily constructed with the following simplified
algorithm~\cite{Newman10}: Every time step a new node is added, with $m$
new edges. Each one of them is connected to an old node, chosen
uniformly at random, with probability $\phi = a/(a+m)$; otherwise, with
the complementary probability $1-\phi = m/(a+m)$, the edge is connected
to a node chose with probability proportional to its in-degree
$q_s(t) - m$. In our simulations we consider LPA networks with minimum
degree $m=2$ and varying $\gamma$, for network sizes ranging from
$N=10^2$ up to $N=10^8$. Topological and spectral properties of LPA
networks are computed averaging over $100$ different network
configurations for each value of $\gamma$ and $N$.

\section{Principal eigenvector localization in linear preferential
  attachment networks}
\label{sec:princ-eigenv-local}

\begin{figure}[t]
  \centering
  \includegraphics[width=0.6\textwidth]{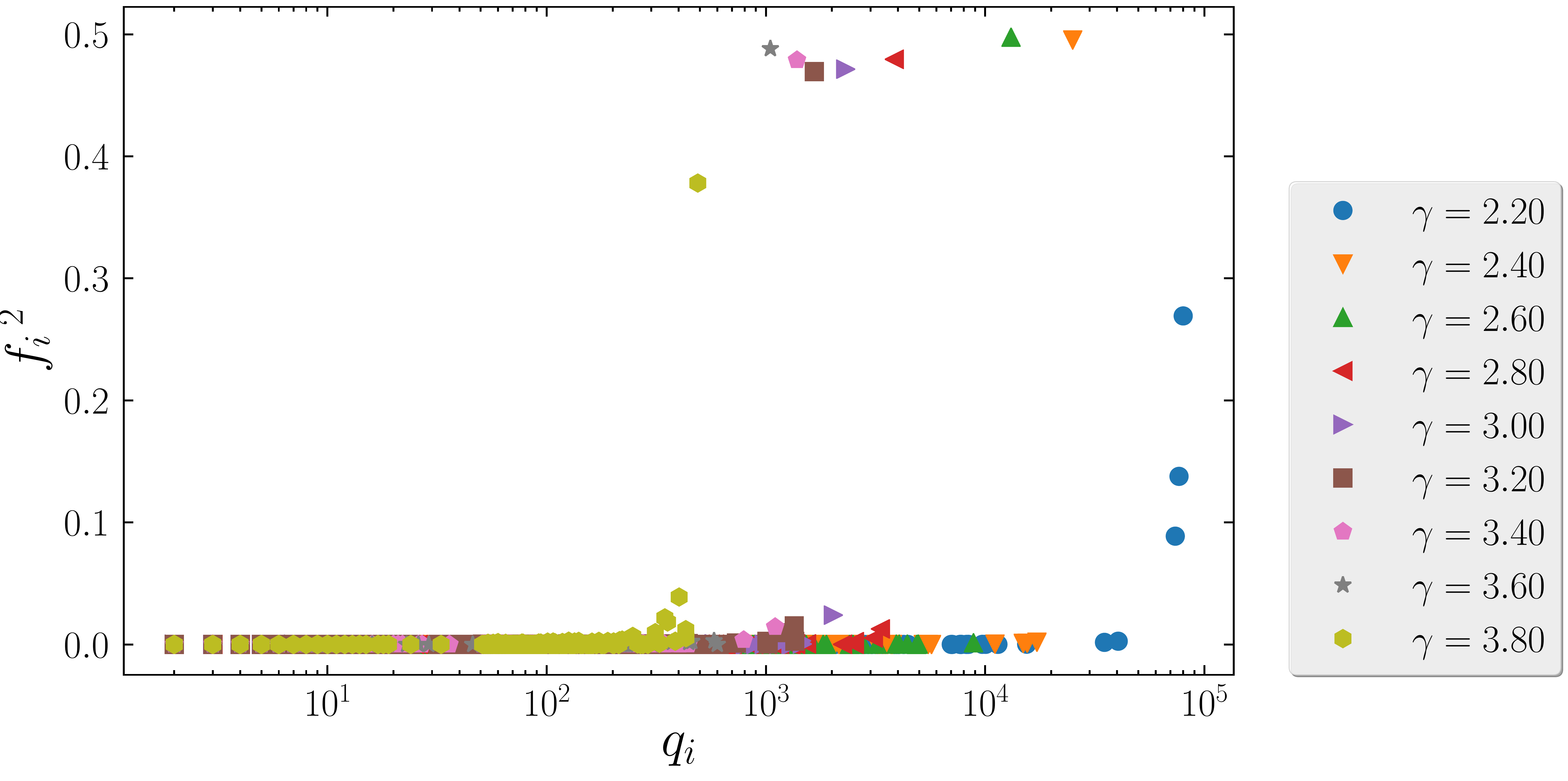}
  \caption{Scatter plot of $f_i^2$ as a function of the degree $q_i$ in
    LPA networks of different degree exponent $\gamma$. Network size
    $N=10^6$.}
  \label{fig:PEV_vs_ki}
\end{figure}

A direct way to observe PEV localization consists in plotting the square
of the components $f_i^2$ as a function of the node degree $q_i$,
Fig.~\ref{fig:PEV_vs_ki}. As we can see from this figure, for all values
of $\gamma$ the component of the PEV associated to the largest values of
$q$ have a macroscopic weight, indicating localization of the PEV in the
hubs. This plot presents evidence of a further difference of LPA
networks with respect to random uncorrelated networks. In this case, and
for $\gamma<5/2$, it is possible to show that the PEV components
approach in static networks the form obtained within the annealed
network approximation~\cite{Boguna09}, which is given
by~\cite{Pastor-Satorras2016}. 
\begin{equation}
  f_i^\mathrm{an} = \frac{q_i}{[N \av{q^2}]^{1/2}}.
\end{equation}
As we can see in Fig.~\ref{fig:PEV_vs_ki}, this linear behavior in not
present in the data from LPA networks, even for small $\gamma$ values.

\section{$K$-core structure in reshuffled linear preferential attachment
  networks}
\label{sec:k-core-structure}

\begin{figure}[t]
  \centering
  \includegraphics[width=0.6\textwidth]{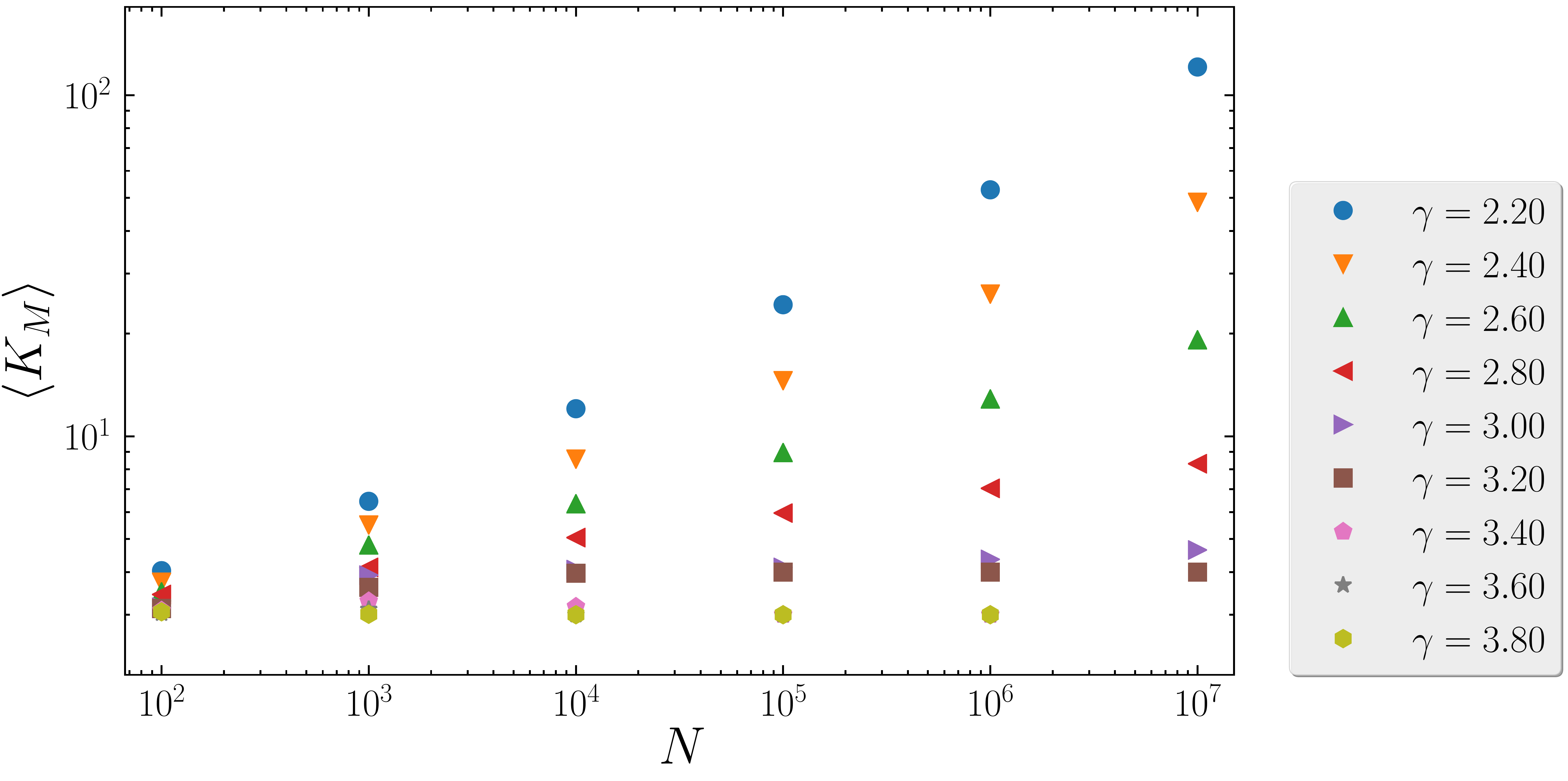}
  \caption{Average maximum core index, $\av{K_M}$ as a function of
    network size for reshuffled LPA networks with different degree
    exponent $\gamma$. Error bars are smaller than symbol sizes.}
  \label{fig:Kcore_randomized}
\end{figure}

The lack of $K$-core structure of LPA networks arises from its peculiar
growing nature, in which nodes with minimum degree $m$ are sequentially
attached to the network. This property is not robust, however, since a
simple reshuffling procedure can destroy it, inducing a non-trivial
$K$-core structure. In Fig.~\ref{fig:Kcore_randomized} we show the
average maximum core index, $\av{K_M}$, as a function of the network
size, computed from LPA networks with different degree exponent, in
which edges have been reshuffled according to the degree preserving edge
rewiring process described in Ref.~\cite{maslov02}. As we can observe,
for $\gamma \geq 3$, the reshuffling process is not able to induce a
substantial $K$-core structure. This occurs because the reshuffling
destroys correlations but uncorrelated networks with $\gamma>3$ have
essentially no $K$-core structure~\cite{Dorogovtsev2006}.
For $\gamma<3$, on the other hand, the
$K$-core structure generated by reshuffling is robust, with an average
maximum core index increasing as a power-law with network
size~\cite{Dorogovtsev2006}.

\begin{figure}[t]
  \centering
  \includegraphics[width=0.6\textwidth]{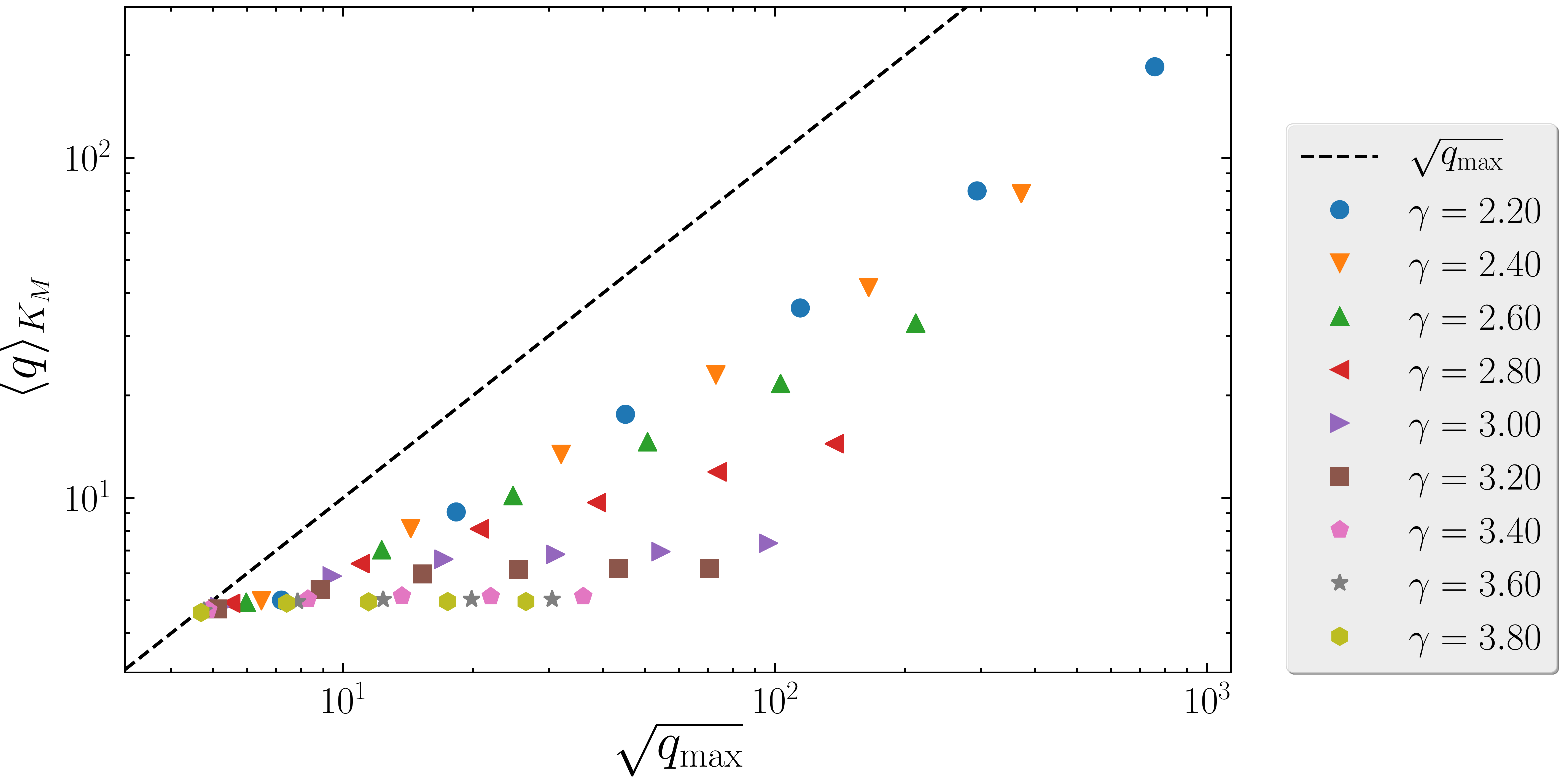}
  \caption{Average maximum core index, $\av{K_M}$ as a function of
    network size for reshuffled LPA networks with different degree
    exponent $\gamma$. Error bars are smaller than symbol sizes.}
  \label{fig:av_qK_randomized}
\end{figure}

The maximum $K$-core resulting from the reshuffling of LPA networks
has an average degree, $\av{q}_{K_M}$, that depends on the maximum
degree $\km$, see Fig.~\ref{fig:av_qK_randomized}. 
However, since the hub has degree much larger than $N^{1/2}$, the network 
does not become uncorrelated even upon randomization
and $\av{q}_{K_M}$ is always smaller than $\sqrt{\km}$.
Hence the properties of the largest eigenpair are always dictated by the hub, 
as in the original LPA networks.

\section{Susceptible-Infected-Susceptible epidemic dynamics}
\label{sec:susc-infect-susc}

The Susceptible-Infected-Susceptible (SIS) model is the simplest model
designed to capture the properties of diseases that do not confer
immunity \cite{Keeling07book}. In the SIS model, individuals can be in
either of two states, susceptible or infected. Susceptible individuals
become infected though a contact with an infected individual at rate
$\beta$, while infected individuals heal spontaneously at rate
$\mu$. As a function of the parameter $\lambda=\beta/\mu$, the model
shows a non-equilibrium phase transition between an active, infected
phase for $\lambda > \lambda_c$, and an inactive, healthy phase for
$\lambda \leq \lambda_c$. Interest is placed on the location of the
so-called epidemic threshold $\lambda_c$, an on its dependence on the
topological properties of the network under consideration
\cite{Pastor-Satorras:2014aa}.

Early theoretical approaches to the SIS dynamics were based on the
so-called Heterogeneous Mean-Field (HMF) theory~\cite{pv01a,Pastor01b},
which neglects both dynamical and topological correlations by replacing
the actual structure of the network, as given by the adjacency matrix,
by an annealed version in which edges are constantly rewired, while
preserving the degree distribution $P(q)$. Within this annealed network
approximation~\cite{dorogovtsev07:_critic_phenom}, a threshold for
uncorrelated networks is obtained of the form
$\lambda_c = \av{q}/\av{q^2}$. An improvement over this approximate
theory is given by Quenched Mean-Field (QMF)
theory~\cite{Chakrabarti_2008}, which, while still neglecting dynamical
correlations, takes into account the full structure of the adjacency
matrix. Within this approximation, the threshold is given by the inverse
of the largest eigenvalue of the adjacency matrix,
$\lambda_c = 1/ \Lambda_M$. Recent and intense activity, based on more
sophisticated approaches~\cite{Castellano2010,
  Ferreira2012,PhysRevLett.111.068701} has shown that on uncorrelated
static networks this result is essentially asymptotically correct.

\begin{figure}[t]
  \centering
  \includegraphics[width=0.6\textwidth]{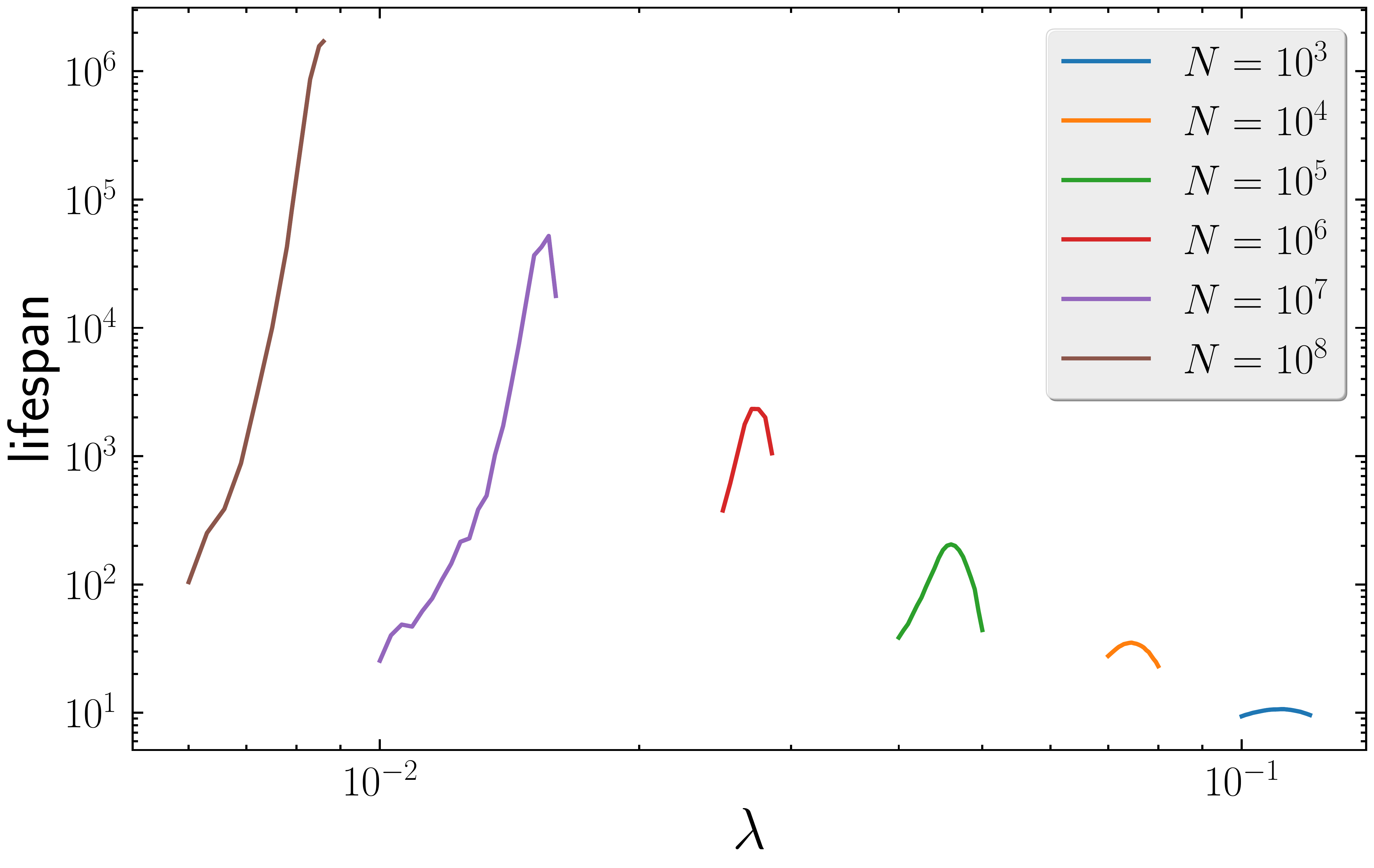}
  \caption{Average lifespan vs spreading rate $\lambda_c$ of SIS
    epidemics starting from a single infected node (the hub) and
    reaching the healthy absorbing state before the coverage reaches the
    threshold value $c=0.5$.  Data are for LPA networks with
    $\gamma=2.6$ and various system size $N$.}
  \label{fig:lifespan}
\end{figure}

In order to determine $\lambda_c$ numerically, we resort to the lifespan
method~\cite{PhysRevLett.111.068701,Mata15}, which is not affected by
the drawbacks that make the consideration of susceptibility
unwieldy~\cite{Ferreira2012}.  Simulations start with only the hub
infected. For each run one keeps track of the coverage, i.e. fraction of
different nodes which have been touched at least once by the
infection. In an infinite network this quantity is vanishing for
$\lambda \leq \lambda_c$, while it tends asymptotically to 1 in the
active region of the phase-diagram.  In finite networks one can set a
threshold $c$ (we choose $c=0.5$) and consider all runs that reach a
coverage larger than $c$ as endemic.  The average lifespan
$\langle T \rangle$ restricted only to nonendemic runs plays the role of
a susceptibility: The
position of the threshold is estimated as the value of $\lambda$ for
which $\langle T \rangle$ reaches a peak, see Fig.~\ref{fig:lifespan}.

\section{Kuramoto synchronization dynamics}
\label{sec:kuram-synchr-dynam}

The Kuramoto model~\cite{RevModPhys.77.137,Rodrigues2016} describes the
dynamics of a collection of weakly coupled nearly identical
oscillators. If they are placed on the nodes of a network with adjacency
matrix $A_{ij}$ the equation of motion reads
\begin{equation}
  \dot{\theta}_i = \omega_i + \kappa \sum_j A_{ij} \sin(\theta_j-\theta_i),
\end{equation}
where $\kappa$ is a coupling constant and $\omega_i$ is a quenched
random variable (natural frequency), whose distribution $g(\omega)$ is
taken here uniform between $-1$ and $1$. In the initial condition the
phases $\theta_i$ are uniformly random between $0$ and $2\pi$.  Defining
the global order parameter as
\begin{equation}
  r = \left| \frac{1}{N}\sum_i e^{-I \theta_i} \right|,
\end{equation}
where $I$ is the imaginary unit, one finds that there is a critical
threshold $\kappa_c$ separating a disordered phase where $r=0$ (in the
thermodynamic limit) from a synchronized phase with $r>0$. A QMF-like
theory for the Kuramoto model~\cite{Restrepo2005} predicts a critical
point $\kappa_c = k_0/\Lambda_M$, where $k_0 = 2/[\pi g(0)] = 4/\pi$,
the last equality holding because of the uniform distribution of natural
frequencies $g(\omega)$.

The value of the critical threshold is numerically determined in
finite networks by computing the susceptibility,
$\chi_K(\kappa) = N(\langle r^2 \rangle- \langle r\rangle^2)$, which
shows a peak for $\kappa = \kappa_c$.
\clearpage

\section{Supplementary Table}
\label{sec:supplementary-table}

\begin{longtable}{||l|c|c|c|c|c|c|c|c|c||}
\hline\hline

Network &      $N$&  $\langle q \rangle$ &  $q_\mathrm{max}$ &  $\chi$
  &     $r$ &  $K_M$ &  $N_{K_M}$ &  $\langle q \rangle_{K_M}$ &
                                                                 $\chi_{K_M}$
  \\ \hline \hline
Social 3               &       32 &                  5.0 &                13 &     0.2 & -0.1194 &      4 &         17 &                        4.7 &        0.0891 \\
Karate club            &       34 &                  4.6 &                17 &     0.7 & -0.4756 &      4 &          9 &                        5.0 &        0.0400 \\
Protein 2              &       53 &                  4.6 &                 8 &     0.2 &  0.2088 &      4 &         29 &                        5.0 &        0.0421 \\
Dolphins               &       62 &                  5.1 &                12 &     0.3 & -0.0436 &      4 &         35 &                        6.1 &        0.0681 \\
Social 1               &       67 &                  4.2 &                11 &     0.2 &  0.1031 &      4 &          5 &                        4.0 &        0.0000 \\
Les Miserables         &       77 &                  6.6 &                36 &     0.8 & -0.1652 &      9 &         12 &                       10.3 &        0.0083 \\
Protein 1              &       95 &                  4.5 &                 7 &     0.1 &  0.1293 &      3 &         83 &                        4.6 &        0.0779 \\
E. Coli, transcription &       97 &                  4.4 &                10 &     0.2 &  0.4116 &      4 &         20 &                        5.4 &        0.0562 \\
Political books        &      105 &                  8.4 &                25 &     0.4 & -0.1279 &      6 &         40 &                        9.1 &        0.1152 \\
David Copperfield      &      112 &                  7.6 &                49 &     0.8 & -0.1293 &      6 &         43 &                        9.5 &        0.2661 \\
College football       &      115 &                 10.7 &                12 &     0.0 &  0.1624 &      8 &        113 &                       10.6 &        0.0067 \\
S 208                  &      122 &                  3.1 &                10 &     0.2 & -0.0020 &      2 &        108 &                        3.2 &        0.1911 \\
High school, 2011      &      126 &                 27.1 &                55 &     0.2 &  0.0829 &     21 &         33 &                       26.7 &        0.0157 \\
Bay Wet                &      128 &                 32.4 &               110 &     0.2 & -0.1117 &     23 &         80 &                       33.2 &        0.0869 \\
Bay Dry                &      128 &                 32.9 &               110 &     0.2 & -0.1044 &     24 &         73 &                       32.7 &        0.0680 \\
Radoslaw Email         &      167 &                 38.9 &               139 &     0.7 & -0.2952 &     36 &         80 &                       51.4 &        0.0559 \\
High school, 2012      &      180 &                 24.7 &                56 &     0.2 &  0.0464 &     18 &         57 &                       23.1 &        0.0246 \\
Little Rock Lake       &      183 &                 26.6 &               105 &     0.6 & -0.2664 &     23 &         64 &                       33.2 &        0.0819 \\
Jazz                   &      198 &                 27.7 &               100 &     0.4 &  0.0202 &     29 &         30 &                       29.0 &        0.0000 \\
S 420                  &      252 &                  3.2 &                14 &     0.2 & -0.0059 &      2 &        226 &                        3.3 &        0.2115 \\
C. Elegans, neural     &      297 &                 14.5 &               134 &     0.8 & -0.1632 &     10 &        118 &                       17.1 &        0.2558 \\
Network Science        &      379 &                  4.8 &                34 &     0.7 & -0.0817 &      8 &          9 &                        8.0 &        0.0000 \\
Dublin                 &      410 &                 13.5 &                50 &     0.4 &  0.2258 &     17 &         32 &                       21.9 &        0.0182 \\
US Air Trasportation   &      500 &                 11.9 &               145 &     3.5 & -0.2679 &     29 &         34 &                       32.5 &        0.0023 \\
S 838                  &      512 &                  3.2 &                22 &     0.3 & -0.0300 &      2 &        462 &                        3.3 &        0.2373 \\
Yeast, transcription   &      662 &                  3.2 &                71 &     3.2 & -0.4098 &      3 &        168 &                        4.9 &        0.3653 \\
URV email              &     1133 &                  9.6 &                71 &     0.9 &  0.0782 &     11 &         12 &                       11.0 &        0.0000 \\
Political blogs        &     1222 &                 27.4 &               351 &     2.0 & -0.2213 &     36 &         55 &                       43.2 &        0.0139 \\
Air traffic            &     1226 &                  3.9 &                34 &     0.9 & -0.0152 &      4 &        106 &                        6.5 &        0.1886 \\
Yeast, protein         &     1458 &                  2.7 &                56 &     1.7 & -0.2095 &      5 &          6 &                        5.0 &        0.0000 \\
Petster, hamster       &     1788 &                 14.0 &               272 &     2.3 & -0.0889 &     20 &        130 &                       31.4 &        0.1399 \\
UC Irvine              &     1893 &                 14.6 &               255 &     2.8 & -0.1880 &     20 &        201 &                       32.1 &        0.1512 \\
Yeast, protein         &     2172 &                  6.0 &               215 &     2.3 & -0.0552 &     10 &         14 &                       11.6 &        0.0093 \\
Japanese               &     2698 &                  5.9 &               725 &    17.3 & -0.2590 &     15 &         52 &                       22.5 &        0.1110 \\
Open flights           &     2905 &                 10.8 &               242 &     4.2 &  0.0489 &     28 &         38 &                       32.8 &        0.0079 \\
GR-QC, 1993-2003       &     4158 &                  6.5 &                81 &     1.8 &  0.6392 &     43 &         44 &                       43.0 &        0.0000 \\
Tennis                 &     4338 &                 37.7 &               451 &     3.2 &  0.0033 &     79 &        308 &                      115.7 &        0.0779 \\
US Power grid          &     4941 &                  2.7 &                19 &     0.5 &  0.0035 &      5 &         12 &                        6.0 &        0.0139 \\
HT09                   &     5352 &                  6.9 &              1287 &    28.0 & -0.4308 &     10 &        319 &                       18.0 &        1.0255 \\
Hep-Th, 1995-1999      &     5835 &                  4.7 &                50 &     0.9 &  0.1852 &     18 &         19 &                       18.0 &        0.0000 \\
Reactome               &     5973 &                 48.8 &               855 &     1.9 &  0.2414 &    176 &        209 &                      197.8 &        0.0036 \\
Jung                   &     6120 &                 16.4 &              5655 &    59.3 & -0.2327 &     65 &        135 &                       76.2 &        0.0359 \\
Gnutella, Aug.8,2002   &     6299 &                  6.6 &                97 &     1.7 &  0.0355 &     10 &        268 &                       17.0 &        0.6928 \\
JDK                    &     6434 &                 16.7 &              5923 &    57.9 & -0.2230 &     65 &        135 &                       76.2 &        0.0359 \\
AS Oregon              &     6474 &                  3.9 &              1458 &    41.4 & -0.1818 &     12 &         20 &                       15.5 &        0.0370 \\
English                &     7377 &                 12.0 &              2568 &    25.8 & -0.2366 &     37 &        112 &                       54.7 &        0.0991 \\
Gnutella, Aug.9,2002   &     8104 &                  6.4 &               102 &     1.6 &  0.0331 &     10 &        315 &                       17.1 &        0.8434 \\
French                 &     8308 &                  5.7 &              1891 &    37.0 & -0.2330 &     17 &         45 &                       23.2 &        0.0490 \\
Hep-Th, 1993-2003      &     8638 &                  5.7 &                65 &     1.3 &  0.2389 &     31 &         32 &                       31.0 &        0.0000 \\
Gnutella, Aug.6,2002   &     8717 &                  7.2 &               115 &     1.0 &  0.0516 &      9 &        175 &                       14.6 &        0.3343 \\
Gnutella, Aug.5,2002   &     8842 &                  7.2 &                88 &     1.1 &  0.0146 &      9 &        238 &                       15.0 &        0.5772 \\
PGP                    &    10680 &                  4.6 &               205 &     3.1 &  0.2382 &     31 &         41 &                       36.5 &        0.0057 \\
Gnutella, Aug.4,2002   &    10876 &                  7.4 &               103 &     0.9 & -0.0132 &      7 &        365 &                       11.8 &        0.3184 \\
Hep-Ph, 1993-2003      &    11204 &                 21.0 &               491 &     5.2 &  0.6295 &    238 &        239 &                      238.0 &        0.0000 \\
Spanish                &    11558 &                  7.4 &              2986 &    60.4 & -0.2819 &     30 &         74 &                       42.7 &        0.0683 \\
DBLP, citations        &    12495 &                  7.9 &               709 &     4.5 & -0.0461 &     12 &        916 &                       21.5 &        0.5223 \\
Spanish                &    12643 &                  8.7 &              5169 &    91.8 & -0.2897 &     30 &         88 &                       44.9 &        0.1029 \\
Cond-Mat, 1995-1999    &    13861 &                  6.4 &               107 &     1.1 &  0.1571 &     17 &         18 &                       17.0 &        0.0000 \\
Astrophysics           &    14845 &                 16.1 &               360 &     1.8 &  0.2277 &     56 &         57 &                       56.0 &        0.0000 \\
Google                 &    15763 &                 18.9 &             11401 &    46.8 & -0.1215 &    102 &        206 &                      106.3 &        0.0129 \\
AstroPhys, 1993-2003   &    17903 &                 22.0 &               504 &     2.0 &  0.2013 &     56 &         57 &                       56.0 &        0.0000 \\
Cond-Mat, 1993-2003    &    21363 &                  8.5 &               279 &     1.6 &  0.1253 &     25 &         26 &                       25.0 &        0.0000 \\
Gnutella, Aug.25,2002  &    22663 &                  4.8 &                66 &     1.2 & -0.1734 &      5 &       6091 &                        8.4 &        0.1481 \\
Internet               &    22963 &                  4.2 &              2390 &    61.0 & -0.1984 &     25 &         70 &                       38.2 &        0.0910 \\
Thesaurus              &    23132 &                 25.7 &              1062 &     3.0 & -0.0477 &     34 &       3644 &                       59.4 &        0.5376 \\
Cora                   &    23166 &                  7.7 &               377 &     2.1 & -0.0553 &     13 &         25 &                       17.4 &        0.0329 \\
Linux, mailing list    &    24567 &                 12.9 &              2989 &    25.5 & -0.1852 &     91 &        157 &                      119.1 &        0.0287 \\
AS Caida               &    26475 &                  4.0 &              2628 &    68.5 & -0.1946 &     22 &         64 &                       33.4 &        0.0688 \\
Gnutella, Aug.24,2002  &    26498 &                  4.9 &               355 &     1.4 & -0.0078 &      5 &       7479 &                        8.8 &        0.3446 \\
Hep-Th, citations      &    27400 &                 25.7 &              2468 &     3.1 & -0.0305 &     37 &         52 &                       44.1 &        0.0068 \\
Cond-Mat,1995-2003     &    27500 &                  8.4 &               202 &     1.6 &  0.1663 &     24 &         25 &                       24.0 &        0.0000 \\
Digg                   &    29652 &                  5.7 &               283 &     3.9 &  0.0027 &      9 &       1339 &                       16.8 &        0.4619 \\
Linux, soft.           &    30817 &                 13.8 &              9338 &    60.6 & -0.1747 &     23 &        439 &                       41.3 &        0.7557 \\
Enron                  &    33696 &                 10.7 &              1383 &    12.3 & -0.1165 &     43 &        275 &                       70.1 &        0.1584 \\
Hep-Ph,citations       &    34401 &                 24.5 &               846 &     1.6 & -0.0064 &     30 &         40 &                       34.4 &        0.0054 \\
Cond-Mat, 1995-2005    &    36423 &                  9.4 &               277 &     2.0 &  0.1776 &     29 &         30 &                       29.0 &        0.0000 \\
Gnutella, Aug.30,2002  &    36646 &                  4.8 &                55 &     1.4 & -0.1038 &      7 &         14 &                        7.0 &        0.0000 \\
Slashdot               &    51083 &                  4.6 &              2915 &    16.9 & -0.0347 &     14 &        736 &                       25.8 &        0.4878 \\
Gnutella, Aug.31,2002  &    62561 &                  4.7 &                95 &     1.5 & -0.0927 &      6 &       1004 &                        9.1 &        0.1253 \\
Facebook               &    63392 &                 25.8 &              1098 &     2.4 &  0.1768 &     52 &        701 &                       88.1 &        0.1538 \\
Epinions               &    75877 &                 10.7 &              3044 &    16.2 & -0.0406 &     67 &        485 &                      113.0 &        0.1608 \\
Slashdot zoo           &    79116 &                 11.8 &              2534 &    11.4 & -0.0746 &     54 &        129 &                       77.8 &        0.0590 \\
Flickr                 &   105722 &                 43.8 &              5425 &     7.0 &  0.2468 &    573 &        574 &                      573.0 &        0.0000 \\
Wikipedia, edits       &   113123 &                 35.8 &             20153 &    18.3 & -0.0651 &    145 &        936 &                      196.7 &        0.1049 \\
Petster, cats          &   148826 &                 73.2 &             80634 &   125.9 & -0.1642 &    419 &       1283 &                      621.7 &        0.0766 \\
Gowalla                &   196591 &                  9.7 &             14730 &    30.7 & -0.0293 &     51 &        185 &                       76.0 &        0.0957 \\
Libimseti              &   220970 &                156.0 &             33389 &     9.5 & -0.1390 &    273 &       4110 &                      475.6 &        0.2285 \\
EU email               &   224832 &                  3.0 &              7636 &   186.7 & -0.1892 &     37 &        292 &                       63.1 &        0.1730 \\
Web Stanford           &   255265 &                 15.2 &             38625 &   132.5 & -0.1156 &     71 &        387 &                      112.2 &        0.3004 \\
Amazon, Mar.2,2003     &   262111 &                  6.9 &               420 &     0.6 & -0.0025 &      6 &        286 &                        6.6 &        0.0389 \\
DBLP, collaborations   &   317080 &                  6.6 &               343 &     2.3 &  0.2665 &    113 &        114 &                      113.0 &        0.0000 \\
Web Notre Dame         &   325729 &                  6.7 &             10721 &    40.9 & -0.0534 &    155 &       1367 &                      157.3 &        0.0591 \\
MathSciNet             &   332689 &                  4.9 &               496 &     2.3 &  0.1030 &     24 &         25 &                       24.0 &        0.0000 \\
CiteSeer               &   365154 &                  9.4 &              1739 &     4.1 & -0.0632 &     15 &       1850 &                       25.4 &        0.3787 \\
Zhishi                 &   372840 &                 12.4 &            127066 &  2243.5 & -0.2825 &    228 &        449 &                      282.7 &        0.0582 \\
Actor coll. net        &   374511 &                 80.2 &              3956 &     4.2 &  0.2260 &    365 &       1178 &                      553.7 &        0.0753 \\
Amazon, Mar.12,2003    &   400727 &                 11.7 &              2747 &     1.6 & -0.0203 &     10 &      27046 &                       13.3 &        0.3788 \\
Amazon, Jun.6,2003     &   403364 &                 12.1 &              2752 &     1.5 & -0.0176 &     10 &      32886 &                       13.4 &        0.4025 \\
Amazon, May5,2003      &   410236 &                 11.9 &              2760 &     1.6 & -0.0169 &     10 &      32632 &                       13.4 &        0.4177 \\
Petster, dogs          &   426485 &                 40.1 &             46503 &    50.3 & -0.0884 &    248 &       1177 &                      386.6 &        0.1198 \\
Road network PA        &  1087562 &                  2.8 &                 9 &     0.1 &  0.1220 &      3 &        916 &                        3.3 &        0.0208 \\
YouTube friend. net.   &  1134890 &                  5.3 &             28754 &    92.9 & -0.0369 &     51 &        845 &                       86.1 &        0.2458 \\
Road network TX        &  1351137 &                  2.8 &                12 &     0.1 &  0.1271 &      3 &       1491 &                        3.4 &        0.0495 \\
AS Skitter             &  1694616 &                 13.1 &             35455 &   109.4 & -0.0814 &    111 &        222 &                      150.0 &        0.0451 \\
Road network CA        &  1957027 &                  2.8 &                12 &     0.1 &  0.1206 &      3 &       4454 &                        3.3 &        0.0268 \\
Wikipedia, pages       &  2070367 &                 40.9 &            230040 &    80.8 & -0.0418 &    208 &        702 &                      283.4 &        0.0895 \\
US Patents             &  3764117 &                  8.8 &               793 &     1.4 &  0.1675 &     64 &        106 &                       76.3 &        0.0079 \\
DBpedia                &  3915921 &                  6.4 &            469692 &  2156.2 & -0.0427 &     20 &         70 &                       27.9 &        0.0620 \\
LiveJournal            &  5189808 &                 18.8 &             15016 &     7.3 &  0.0394 &    374 &        415 &                      408.9 &        0.0006 \\
\hline\hline

  
  \caption{Topological properties of the real networks considered:
    Network size $N$; average degree $\av{q}$; maximum degree
    $q_\mathrm{max}$; heterogeneity parameter
    $\chi = \av{q^2}/\av{q}^2 -1$; degree correlations as measured by
    the Pearson coefficient $r$~\cite{assortative}; maximum core index
    $K_M$; size of the maximum core $N_{K_M}$; average internal degree
    of the maximum core $\langle q \rangle_{K_M}$; heterogeneity
    parameter of the maximum core
    $\chi_{K_M} = \av{q^2}_{K_M}/\av{q}^2_{K_M} -1$.}
\label{tab:myfirstlongtable}
\end{longtable}

\clearpage

\bibliography{Resubmit_Clean}

\end{document}